\documentclass[twocolumn]{aa} 
\usepackage{longtable}
\usepackage{graphicx}
\usepackage{txfonts}
\usepackage{csvsimple}
\usepackage{multirow}
\usepackage{rotating}
\usepackage{amsmath}
\usepackage{longtable}
\usepackage{booktabs}
\usepackage{ltablex}
\usepackage{epsfig}
\usepackage{threeparttable}
\usepackage{url}
\usepackage{xcolor}
\usepackage{color}
\usepackage{hyperref}
\usepackage{soul}

\def\cm2{\rm \ cm$^{-2}$}

\def\deg{\mbox{$^{\circ}$}}

\def\kp{${\rm Kp}$}

\def\deg{\mbox{$^{\circ}$}}

\begin{document}

   \title{In-flight calibration of the {\em Lobster Eye Imager for Astronomy}}


   \author{Huaqing Cheng
          \inst{1}
          \and
          Hai-Wu Pan\inst{1}
          \and
          Yuan Liu\inst{1}\fnmsep\thanks{Corresponding: liuyuan@nao.cas.cn}
          \and
          Jingwei Hu\inst{1}
          \and
          Haonan Yang\inst{1}
          \and
          Donghua Zhao\inst{1}
          \and
          Zhixing Ling\inst{1,2}
          \and
          He-Yang Liu\inst{1}
          \and
          Yifan Chen\inst{3}
          \and
          Xiaojin Sun\inst{3}
          \and
          Longhui Li\inst{4}
          \and
          Ge Jin\inst{4}
          \and
          Chen Zhang\inst{1,2}
          \and
          Shuang-Nan Zhang\inst{5,2}
          \and
          Weimin Yuan\inst{1,2}
          }

   \institute{Key Laboratory of Space Astronomy and Technology, National Astronomical Observatories, Chinese Academy of Sciences, Datun Road 20A, Chaoyang District, Beijing 100101, China
         \and
             School of Astronomy and Space Science, University of Chinese Academy of Sciences, Beijing 100049, China
        \and
            Shanghai Institute of Technical Physics, Chinese Academy of Sciences, Yutian Road 500, Shanghai 200083, China
        \and
            North Night Vision Technology Co., LTD, Nanjing, 211106, China
        \and
            State Key Laboratory of Particle Astrophysics, Institute of High Energy Physics, Chinese Academy of Sciences, Beijing 100049, China}

   \date{Received September XX, XXXX; accepted March XX, XXXX}


  \abstract
{The Lobster Eye Imager for Astronomy ({\em LEIA}), as a pathfinder of the Wide-field X-ray Telescope (WXT) onboard the \textit{Einstein Probe} (\textit{EP}) satellite, is the first lobster-eye focusing X-ray telescope with a considerably large field-of-view (FoV) ever flown.
During the two and half years of operations, a series of calibration observations were performed, to fully characterize its in-orbit performance and calibrate the instrumental properties.
In this paper, we present the results of the in-flight calibration campaign of {\em LEIA}, focusing on the properties of the point spread function (PSF), source positional accuracy, effective area, energy response and the instrumental background.
The calibration sources used are the Crab nebula, Sco X-1 and Cassiopeia A supernova remnant. 
Specifically, it is found that the spatial resolution remains almost unchanged compared to the pre-launch values, ranging from $3.6'$--$9.3'$ with a median of $5.9'$. 
The post-calibration source positional accuracy is found to be $\sim$2$'$ (at the 90\% confidence level).
The Crab spectra can be well reproduced by the absorbed power-law model with the best-fit parameters in large agreement with the literature values, indicating that the in-orbit effective area is overall consistent with the model predictions and ground measurements.
The effective area exhibits a systematic of $\lesssim$10$\%$ (at the 68\% C.L.), and a mild deterioration of $\sim15\%$ at the lower energy end after one year of operation.
The Cas A spectral analysis shows that the energy scale and spectral resolution of the detectors are generally consistent with ground values.
The instrumental background is found to be largely consistent among the four detectors, with strong modulations by the geomagnetic activity and the spectrum qualitatively consistent with our previous simulations.
These instrumental performances well meet the design requirements.
This work paves the way for the in-orbit calibration of the {\it EP}-WXT.
} 

   {}
   {}
   {}
   {}

               
   \maketitle
%

\section{Introduction}

Wide-field X-ray focusing imaging telescopes set to play a vital role in monitoring the X-ray sky, enabling unprecedented studies of transient phenomena and high energy astrophysical processes. 
The Lobster Eye Imager for Astronomy \citep[{\em LEIA},][]{2023LingZXRAA}, serving as a pathfinder for the \textit{Einstein Probe}\footnote{The EP mission is an interdisciplinary X-ray Observatory dedicated for time-domain astronomy and high-energy astrophysics, led by the Chinese Academy of Sciences (CAS) with collaborations from the European Space Agency (ESA), the Max-Planck Institute for Extraterrestrial Physics (MPE), and the France Space Agency (CNES).} \citep[\textit{EP},][]{2016SSRv..202..235Y,2018SPIE10699E..25Y,Yuan2022, 2025EP}, represents a breakthrough in wide-field X-ray monitoring.
As a qualification model (QM) of one of the twelve identical modules of \textit{EP}'s Wide-field X-ray Telescope (WXT), \textit{LEIA} is featured with the novel lobster-eye micro-pore optics \citep[MPO, e.g.][]{Angel1979,Fraser1992, Fraser1993,Willingale1998,Willingale2016}, marking the first of such kind with a considerable large field of view (FoV, $18.6\deg\times18.6\deg$) ever flown.
Moreover, it is also the first application of the large-format scientific complementary metal-oxide semiconductor (CMOS) sensors \citep[e.g.][]{2022WangWXa,2022WangWXb,Wu2022,Wu2023pasp1,Wu2023pasp2,Wu2023nima,ChenMX2024,LiuMJ2025} as X-ray detector in orbit.
The \textit{LEIA} experiment was first proposed in 2019 by the \textit{EP} team and successfully launched on July 27, 2022 as one of the six scientific payloads onboard the SATech-01 satellite developed by the Chinese Academy of Sciences (CAS). 
The goals of the {\em LEIA} experiment are to demonstrate the technologies and in-orbit performance of the MPO and CMOS utilized on \textit{EP}-WXT in advance, as well as to explore the in-orbit calibration procedures for such wide-field lobster-eye X-ray telescope, and to optimize the working configurations of the instrumental parameters and conditions in operation \citep{2022ZhangChenApJL, 2023LingZXRAA}. 

The {\em LEIA} instrument had been in operations for over two and half years before its decommissioning in February 2025, during which its primary goals have been achieved. 
The first wide-field X-ray focusing images were obtained during its commissioning phase, showcasing an excellent in-orbit performance of the instrument after its launch \citep{2022ZhangChenApJL}.
In addition, some interesting scientific discoveries have also been made. 
For instance, the prompt X-ray emission of the second brightest Gamma-ray burst GRB 230307A was caught by \textit{LEIA} precisely at the trigger time.
Analysis of these data revealed compelling evidence for the emergence of a magnetar central engine, marking the first time such a phenomenon has been observed during the prompt emission of a GRB \citep{2025SunNSR}.
The longest-lasting and most energetic stellar X-ray flare from a nearby K-type giant star HD 251108 was captured by \textit{LEIA} on November 7, 2022 \citep{2025MaoSF}.
By combining the X-ray data of \textit{LEIA} with other telescopes and instruments, \citet{Yang2025BeXRB} investigated the new giant outburst of the Be X-ray binary RX J0520.5-6932 occurred in March 2024.
Observations of the Virgo Cluster validated the instrument's capabilities for extended source imaging and spectral resolution \citep{2024FengVirgo}.
Furthermore, \textit{LEIA} conducted a densely daily monitoring campaign of the Large Magellanic Cloud (LMC) region for over two years, which systematically documents the long-term activities of bright X-ray binaries such as LMC X-1, X-2, X-3, and X-4 (Yang et al. in preparation).

On the other hand, the in-flight calibration procedures and data analysis methods have been successfully validated via a series of calibration observations conducted during different stages of the operation phase.
Comprehensive pre-launch instrumental characterization on different levels of devices, assemblies and the complete module, culminated in an end-to-end calibration campaign carried out at the 100-m X-ray Facility \citep[100XF,][]{2023WangYusa} at the Institute of High Energy Physics (IHEP) of CAS in November 2021. 
The results from these efforts \citep{Cheng2024a} were incorporated into the initial calibration database (CALDB), which has been applied to the post-launch data analysis. 
While sharing a variety of commonalities with the in-orbit calibration of Wolter-I type telescope, such as the standard calibration X-ray sources and common spectral/timing analysis methods, \textit{LEIA}'s calibration paradigm addresses unique challenges induced by its innovative design. 
The main concern is how to ensure a largely uniform source positional accuracy, at the precision of a few arc-minutes, across essentially the entire FoV of over 300 square degrees. 
Also, the uniform PSF characteristics and effective area, as theoretically predicted by lobster-eye optics \citep[e.g.][]{Angel1979, Fraser1992} and preliminarily verified through ground calibration experiments \citep{Cheng2024a} need to be verified as well. 
Additional complexities arise from the pioneering use of scientific CMOS detectors in X-ray astronomy \citep{2023LingZXRAA}, entailing long-term monitoring of energy response stability. 
These requirements necessitated developing specialized calibration protocols, which now serve as the heritage for the \textit{EP} mission. 
The demonstrated success of \textit{LEIA}'s calibration paradigm confirms the viability of lobster-eye optics for future time-domain X-ray missions while establishing new standards for wide-field space instrumentation.

In this paper, we present the results of the in-flight calibration of \textit{LEIA} in details, including the properties of the point spread function (PSF), source positional accuracy, effective area, energy response of the CMOS sensors and the instrumental background.
Due to the presence of the localized PSF misalignment within three confined regions of the two FoV quadrants subtended by CMOS 1 and CMOS 2 \citep[see Section 3.1 and Section 4.3 in][for more details]{Cheng2024a}, the in-orbit calibration of \textit{LEIA} focused on the other two FoV quadrants corresponding to CMOS 3 and CMOS 4, the performance of which are more representative of the \textit{EP}-WXT flight model (FM) modules \citep{2025ChengWXTcalib}.
Except for the background section, the results presented in this paper all derive from observations of CMOS 3 and CMOS 4.

The structure of the paper is as follows. In Section \ref{sec:calib_procedure}, we briefly review the calibration campaign of {\em LEIA} after the launch, including the calibration stages, calibration sources and data reduction procedures of the calibration data. 
We present the calibration results of PSF in Section \ref{sec:psf}, source positional accuracy in Section \ref{sec:alignment_calibration}, effective area in Section \ref{sec:effarea}, energy response in Section \ref{sec:cas_a_data_analysis} and instrumental background in Section \ref{sec:background}. The summary and conclusion are given in Section \ref{sec:summary}.

\section{Overview of the in-flight calibration campaign}
\label{sec:calib_procedure}

\subsection{Calibration stages}
\label{sec:calib_stage}

\begin{table*}[hbtp]
\caption{The basic information of the in-flight calibration campaign of \textit{LEIA}.}             
\label{table:calibplan}      
\centering                          
\begin{tabular}{ccccccc}        
\hline                
\hline
Round & Target & Observation date & Calibration term related & Calibrated detectors & Test points per detector \\    
\hline
1 & Cas A & 2022.8 & Gain, Energy resolution & 3, 4 & 1 \\
  & Sco X-1 & 2022.8 & PSF & 3, 4 & 1 \\
  & Crab & 2022.10-2023.2 & PSF, positional, Effective area & 1, 2, 3, 4 & 121 \\
\hline
2 & Cas A & 2023.2 & Gain, Energy resolution & 3, 4 & 1 \\
  & Crab & 2023.3 & Effective area & 1, 2, 3, 4 & 9 \\
\hline
3 & Cas A & 2023.10 & Gain, Energy resolution & 3, 4 & 1 \\
  & Crab & 2023.10 & Effective area & 3, 4 & 1 \\
\hline
\hline                                   
\end{tabular}
\end{table*}

The basic information of the in-flight calibration campaign of \textit{LEIA} is summarized in Table \ref{table:calibplan}, including the chronological stages, observational targets, calibration parameters, calibrated detectors, and spatial sampling pattern.
The calibration campaign of \textit{LEIA} can be divided mainly into three stages. 
Commencing from the observation of Cassiopeia A supernova remnant (Cas A) in August 2022 and culminating in February 2023 with observations of the Crab nebula (Crab), the first calibration stage lasted for about seven months.
As the most extended calibration phase, we implemented a thorough and comprehensive characterization of the instrumental performances and properties after the launch, including the PSF, effective area and energy response of the CMOS detectors.
More importantly, the source localization calibration was accomplished in this stage.

The second and third calibration stages aim to characterize the monthly and yearly evolution of the instrumental properties, which are presumed to be stable based on ground tests and simulations. 
The second calibration was carried out in February and March 2023, and the third calibration in October 2023.
Specifically, the Crab and Cas A were mainly observed at the center of the CMOS detectors during these stages, to assess the potential degradations in the effective area and detector response. 

\subsection{Calibration sources}
\label{sec:calib_sources}

For \textit{LEIA}, the Crab serves as the principal calibration source for the effective area, PSF characterization, and source localization.
In practice, it is one of the standard celestial calibration sources on the effective area for X-ray telescopes by virtue of its relative stability, brightness and featureless spectrum \citep[e.g.][]{Toor_Seward_1974,Kirsch2005,Weisskopf2010Crab,2015Mason_NuSTAR_calibration}, though it is somewhat too bright for reliable pileup-corrections for most of the focusing X-ray instruments employing CCD detectors. 
In the meantime, while the nebula exhibits an angular extent of $\sim120''\times100''$,  it can be treated as a point source and used for PSF calibration due to the fact that the X-ray emission is predominantly contributed by the central pulsar \citep[e.g.][]{2015Mason_NuSTAR_calibration, Nakajima_Hiroshi_Hitomi2018}.
To ensure consistency with ground calibration experiments, we implemented a sampling of the PSF, effective area and positional offset using a grid of $22\times22$ points corresponding to 484 incident directions within the entire FoV.
We note that during the first two months when the Crab nebula was invisible due to Solar angle constraint, the brightest persistent X-ray source Scorpius X-1 \citep[Sco X-1,][]{1962GiacconiScoX1} provided additional PSF performance validation, leveraging CMOS's small-size pixel and fast readout speed to mitigate pileup effects.

The energy response calibration of the CMOS detectors, including the Energy-Channel (EC) relation (i.e. gain coefficient) and spectral resolution, necessitated bright X-ray sources due to the instrument's limited photon collecting capability \citep[the maximum effective area is $\sim$3$~{\rm cm^2}$ at $\sim$1 keV,][]{Cheng2024a}. 
We selected Cas A for this purpose, leveraging its relatively high X-ray brightness ($F_{\rm 0.5-4~keV}\sim10^{-9}~{\rm erg~s^{-1}~cm^{-2}}$) and strong emission lines (particularly Si He-$\alpha$ at $\sim$1.86 keV and S He-$\alpha$ at $\sim$2.45 keV) to characterize spectral performance across the 0.5--4 keV band. 

\subsection{Data reduction}
\label{sec:data_reduction}

The calibration observations were performed in photon-counting (PC) mode. For each observation, the X-ray events were processed and calibrated using \texttt{WXTDAS}, the dedicated data reduction software developed for the \textit{EP} mission (Liu et al., in preparation). The event energies, originally recorded in the Pulse-Height-Amplitude (PHA), were corrected and transformed into the unit of PHA-Invariant (PI) using the bias and gain values stored in the first version of CALDB, which was constructed based on ground calibration experiments \citep{Cheng2024a}. 
Additionally, bad and flaring pixels were flagged during the data processing. 
The positions of the events were projected into celestial coordinates (J2000), using the detector-to-celestial transform matrix stored in the CALDB. 
Single-, double-, triple-, and quadruple- events without anomalous flags were selected to generate a clean event file.
Based on this file, an observational image and exposure map were generated. A source catalog comprising a list of the high-significance sources \citep[exceeding $5\sigma$, estimated by employing the Li-Ma method,][]{LiMa1983} was derived based on a point-source detection algorithm. Light curves and spectra of both the source and background were extracted. 
For the source, a circular region with a radius of $9.2'$ was used, which encloses more than 90\% of the focal spot photons. 
For the background, an annular region centering on the source position with inner and outer radii of $18.4'$ and $36.8'$, respectively, was applied. To perform spectral analysis, ancillary response files (ARFs) and response matrix files (RMFs) were generated. For the Crab observations, the source spectra were rebinned to ensure a minimum of 25 photons per energy bin using the {\em grppha} tool (version 3.1.0)\footnote{\url{https://heasarc.gsfc.nasa.gov/docs/journal/grppha4.html}}. 

\section{Point spread function}
\label{sec:psf}

\begin{figure*}[!htbp]
    \centering
    \includegraphics[width=0.48\textwidth]{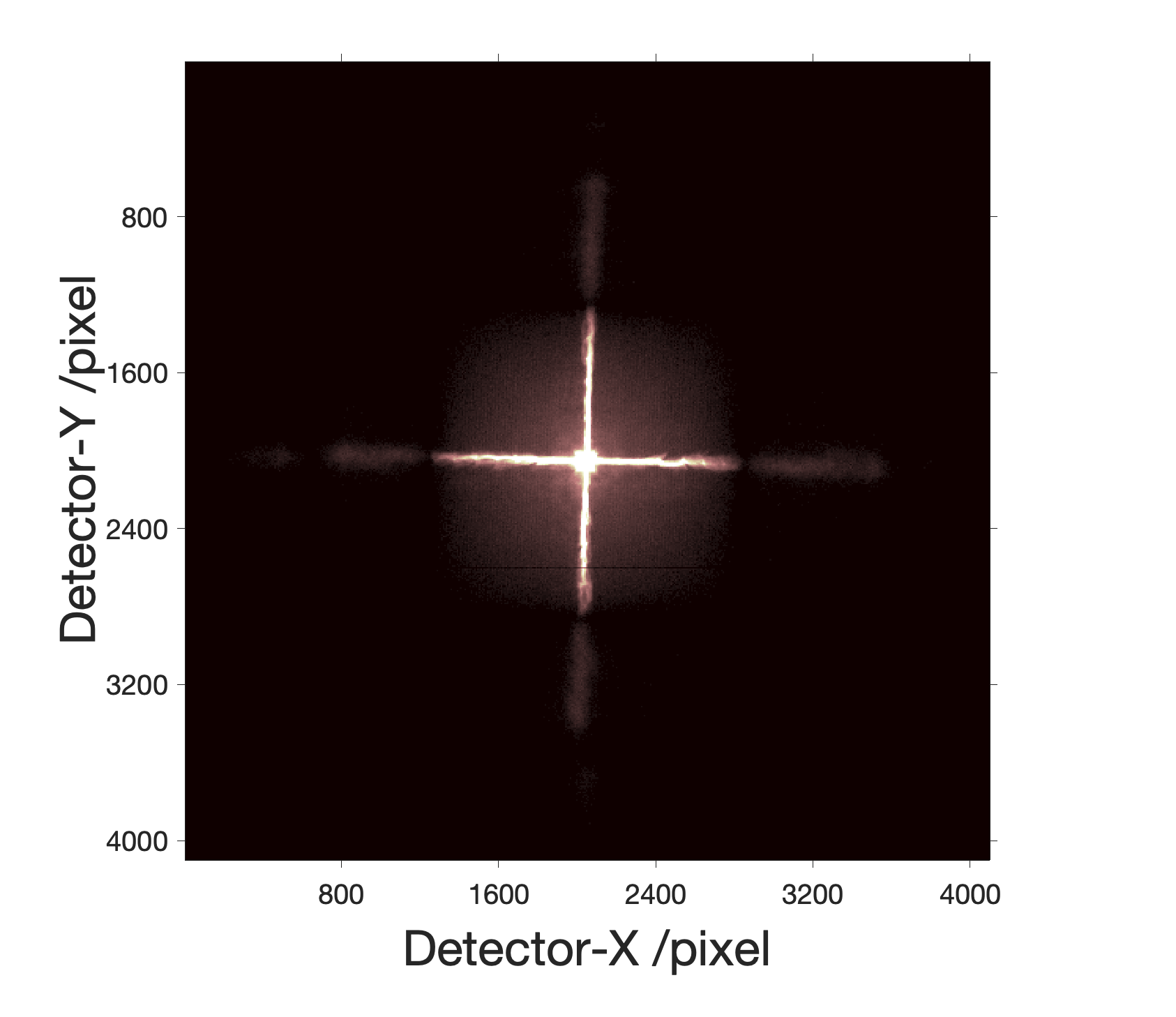}
    \includegraphics[width=0.48\textwidth]{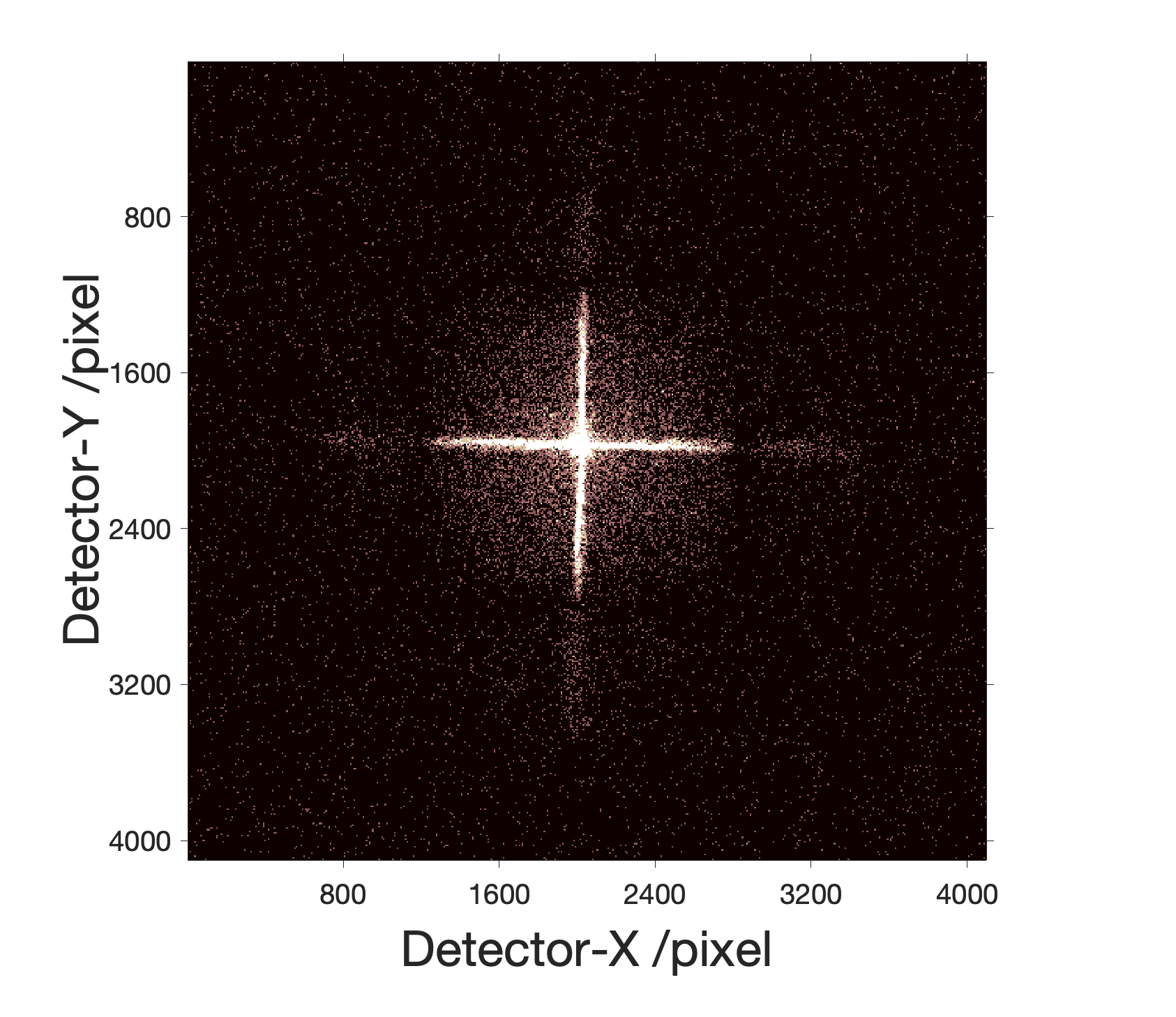}
    \caption{({\em Left} panel): 
    On-ground PSF measured at the center of the CMOS 3 detector at XIB/NAOC using an X-ray spectrum peaking at $\sim$2.5 keV \citep[adapted from Figure 2 of][]{2022ZhangChenApJL}.
    ({\em Right} panel): Observed image of Sco X-1 on the CMOS 3 of {\em LEIA}, taking photons within 2--3 keV energy band.
    Both images are displayed on logarithmic color scales, binned $8\times8$, and shown in detector coordinates.}
    \label{fig:psf_scox1}
\end{figure*}

\begin{figure*}[!htbp]
    \centering
    \includegraphics[width=0.8\textwidth]{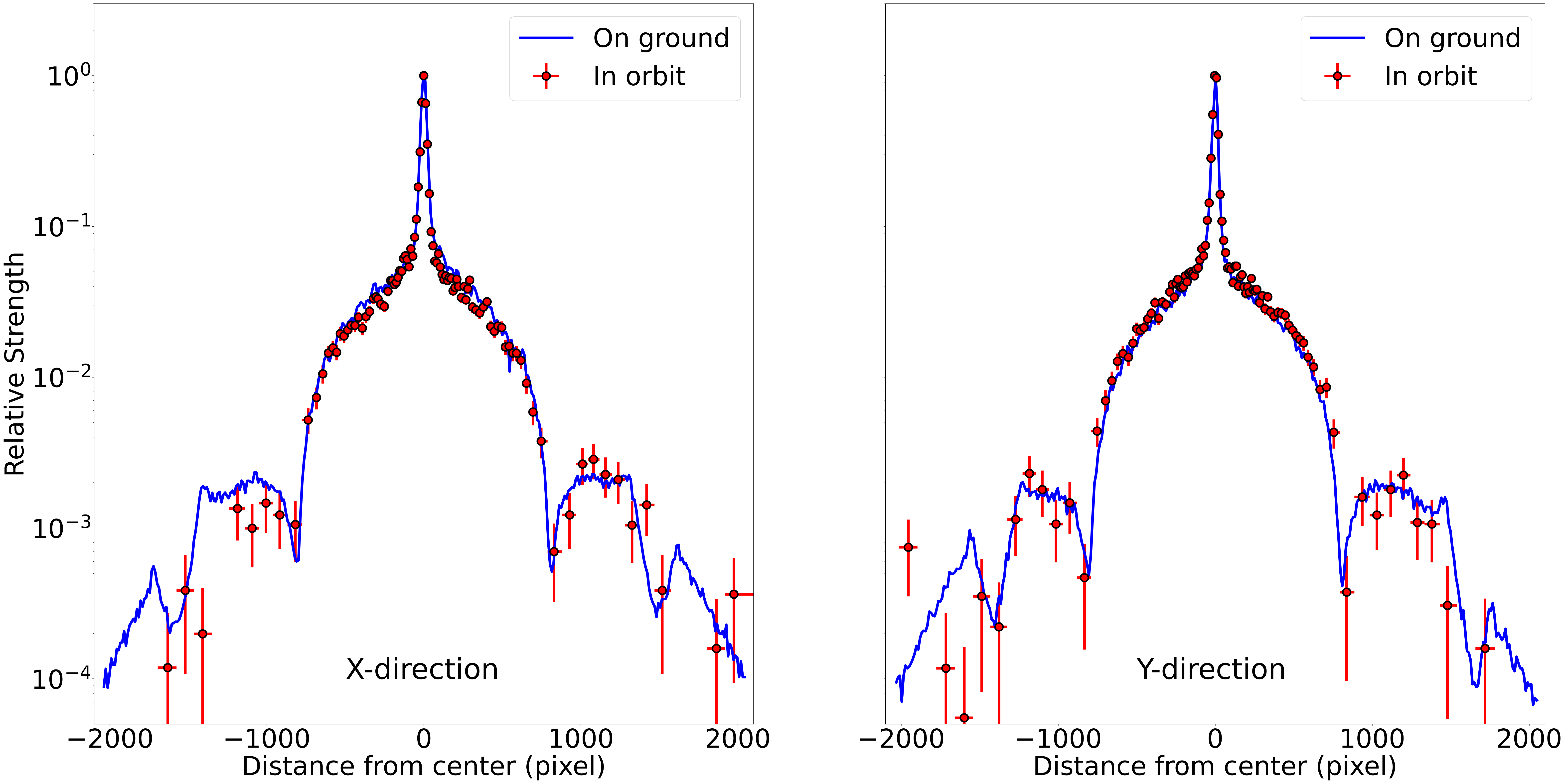}
    \caption{A comparison of the projection of the PSF along the detector X-axis ({\em Left} panel) and the Y-axis ({\em Right} panel). The blue lines are from the ground PSF, and the red symbols are from the Sco X-1 data (the background has been subtracted).}
    \label{fig:scox1-psf-projection}
\end{figure*}

\begin{figure*}[!htbp]
    \centering
    \includegraphics[width=0.5\textwidth]{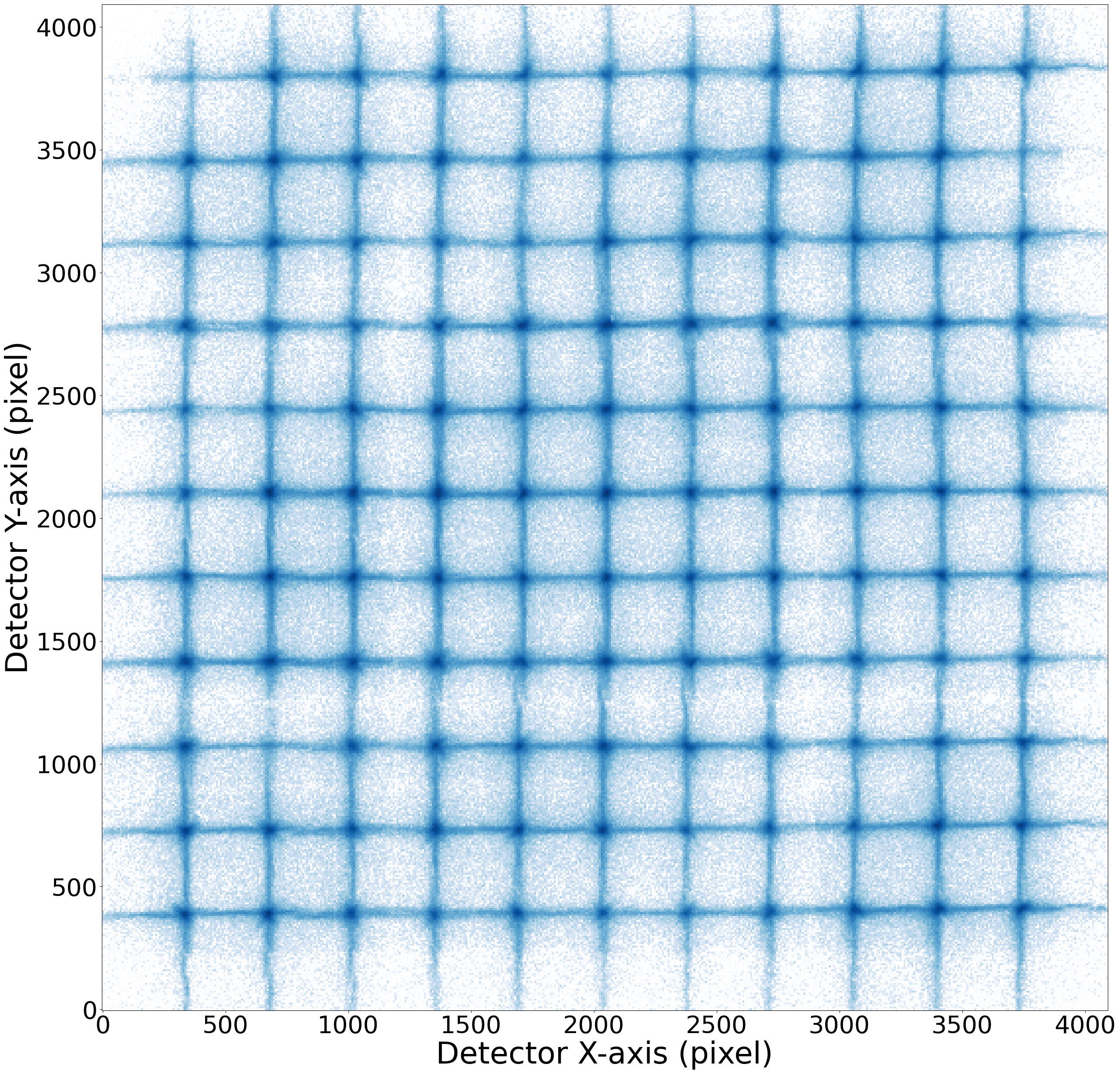}
    \caption{A mosaic of images of Crab observations using a grid of $11\times11$ points corresponding to 121 incident directions on CMOS 4.}
    \label{fig:crabscan}
\end{figure*}

\begin{figure*}[!htbp]
    \centering
    \includegraphics[width=\textwidth]{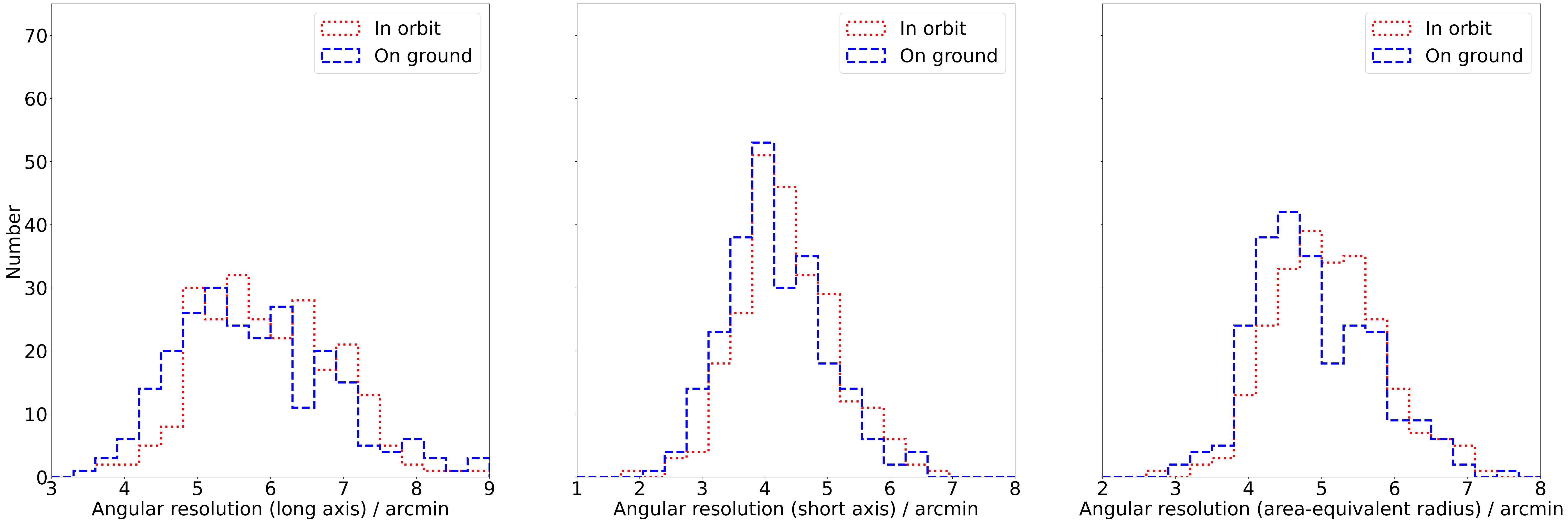}
    \caption{The distributions of the three measures of the PSF FWHM obtained from 242 Crab observations of CMOS 3 and 4. From {\em Left} to {\em Right} panels: the lengths of the long axis, short axis and area-equivalent radius. The result from the ground calibration campaign conducted at 100-m X-ray facility is also plotted for comparison.}
    \label{fig:fwhm_variation_3types}
\end{figure*}

To demonstrate the imaging quality of {\em LEIA}, the brightest non-transient X-ray source Sco X-1 was observed on August 26, 2022 at CMOS 3, shortly after the launch (note that the Crab was invisible during August and September).
It is worth noting that the small pixel size (15 $\mu$m $\times$ 15 $\mu$m) and the fast readout speed of the CMOS detectors ensure that even for Sco X-1 the pileup effect is negligible. In general, for a source with brightness of $\sim25~{\rm Crab}$ / 1000 $\rm{counts~s^{-1}}$, the pileup fraction is less than $1\%$ \citep{2022ZhangChenApJL}. 
The observational image of Sco X-1 at the center of CMOS 3 is presented in Figure \ref{fig:psf_scox1}.
For comparison, the on-ground point-source PSF obtained at the X-ray Imaging Beamline of the National Astronomical observatory of CAS \citep[XIB/NAOC,][]{2012zhangSPIE} is also plotted. 
Following the analysis procedure implemented to the ground calibration data, the $3\times3$ binned image is analyzed under the detector coordinate system (i.e. RAWX and RAWY).
An elliptical function is employed to fit the half-height contour of the PSF focal spot region, in order to derive the three measures of the Full Width at Half Maximum (FWHM) of the PSF, i.e. the lengths of the long axis, short axis and area-equivalent radius \citep[for a detailed description of the PSF analysis, see Section 3.1 of][]{Cheng2024a}.
Specifically, the best-fit ellipse shows major and minor axes of $3.5'$ and $2.7'$, respectively, representing a modest improvement over ground-based measurements ($4.1'$ and $3.3'$).
This agrees with our expectations given the $\sim0.5'$ spatial extent of the X-ray source at XIB/NAOC. 
Furthermore, it is found that the projected PSF profile exhibits a good consistency along the directions of the two cruciform arms,
as presented in Figure \ref{fig:scox1-psf-projection} (note that the background has been subtracted for the observed Sco X-1 image).

For the focal plane mapping of the PSF, the mosaic image of the 121 Crab observations conducted on CMOS 4 is presented in Figure \ref{fig:crabscan}, as an example.
It is found that the cruciform PSFs are largely uniform in shapes across the FoV as predicted by the lobster-eye optics, and are well aligned with each other as shown in ground calibrations.
We note that the fainter focal spots of a few sampled directions are mainly a result of shorter exposure time in these observations (< 200 seconds).
In Figure \ref{fig:fwhm_variation_3types}, we present the distribution of the three FWHM measures of the sampled PSFs obtained on CMOS 3 and CMOS 4, with the ground measurement obtained at 100-m X-ray facility over-plotted for comparison.
It is found that the FWHM of the in-orbit PSF ranges from $3.6'$--$9.3'$ (with a median of $5.9'$) and $1.9'$--$6.8'$ (with a median of $4.3'$) for the long and short axial directions, respectively.
The former is adopted as the indicator of spatial resolution.
All three measures of the PSF FWHM are consistent with ground measurements.
In this sense, we conclude that no noticeable degradation of the imaging quality is observed after the launch.

\section{Source positional accuracy}
\label{sec:alignment_calibration}

\begin{figure*}[!htbp]
    \centering
    \includegraphics[width=0.45\textwidth]{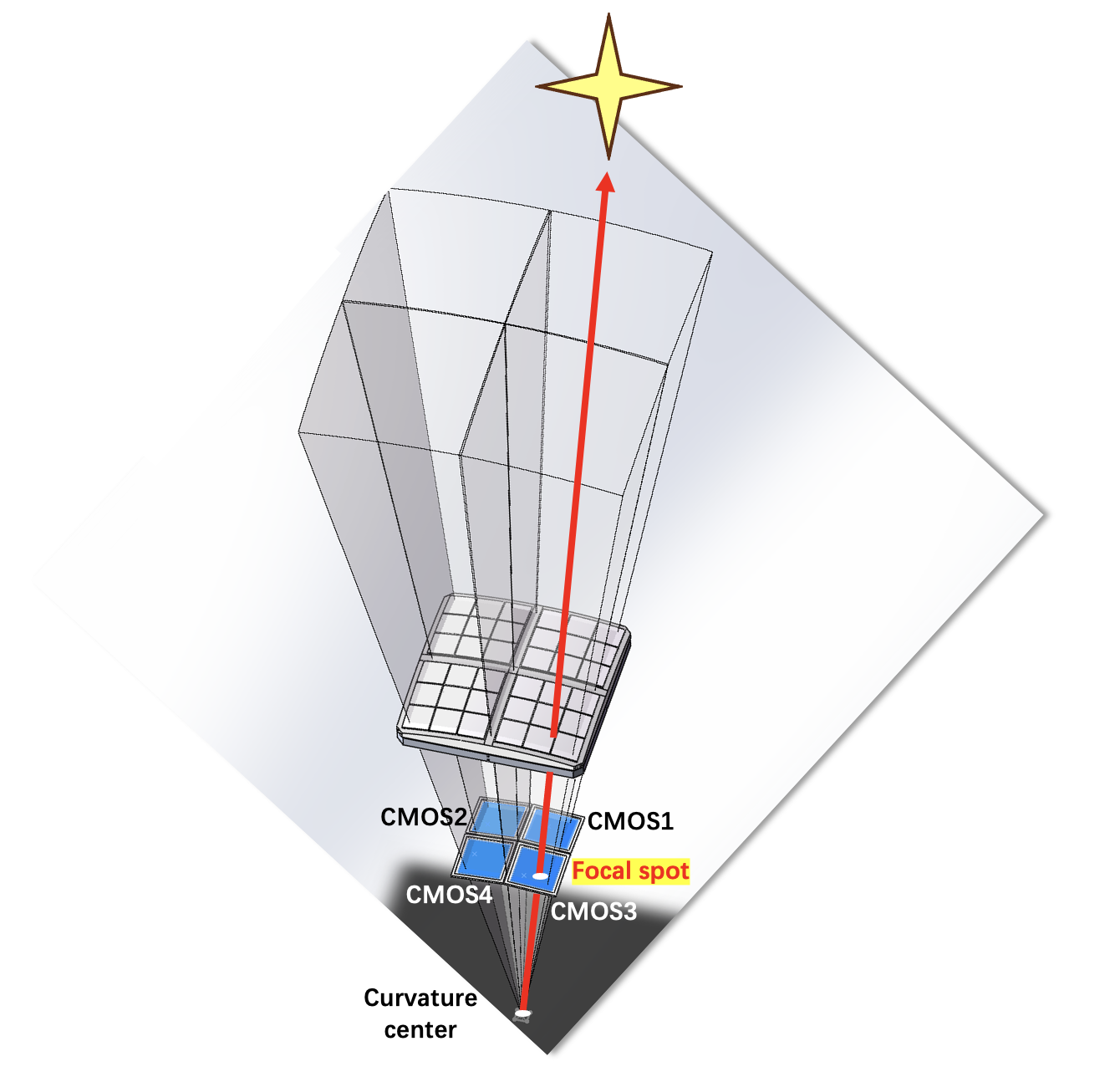}
    \includegraphics[width=0.45\textwidth]{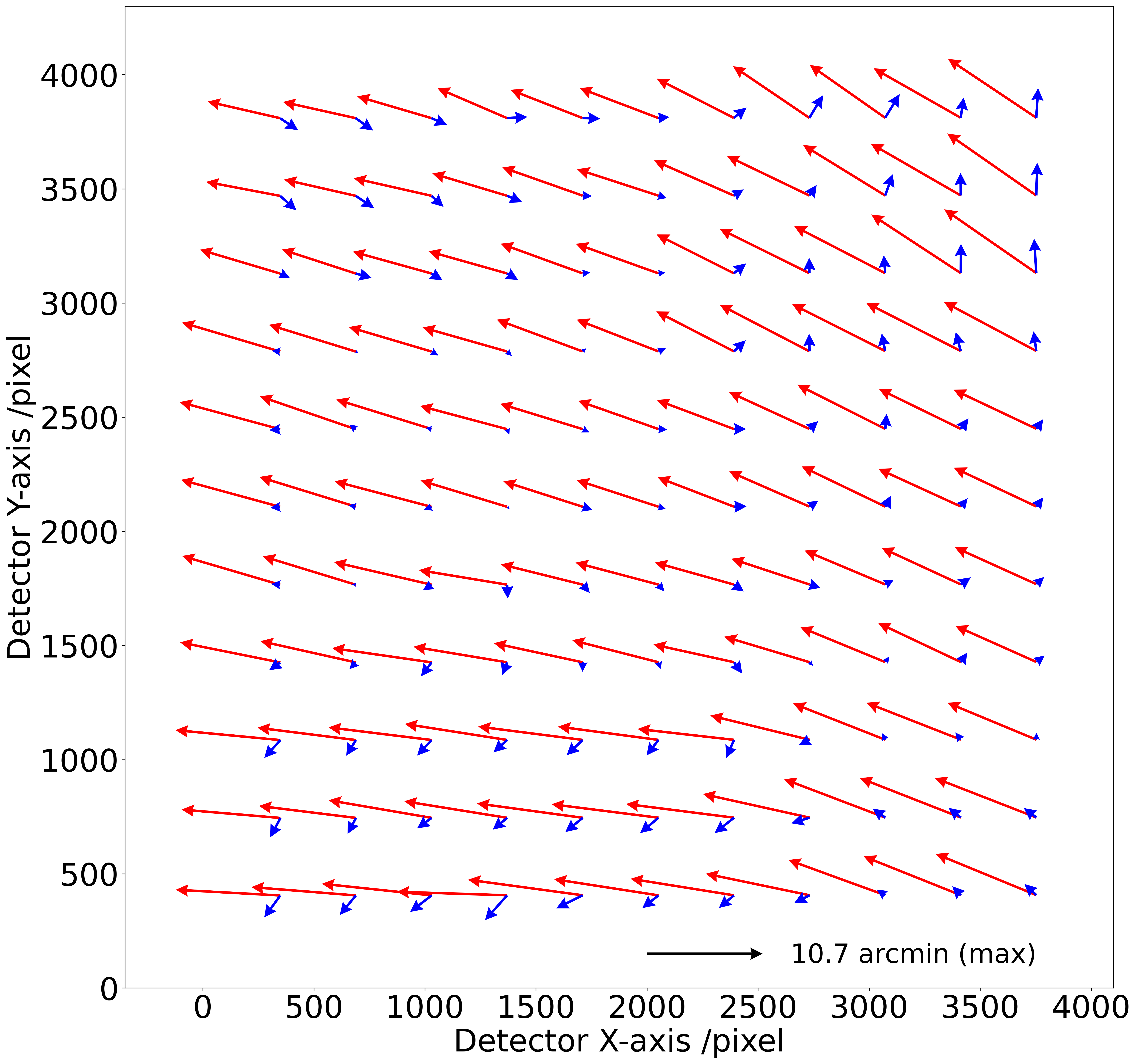}
    \caption{({\em Left} panel): Schematic diagram of the \textit{LEIA} focusing system, focal plane detector array, and field of view (FoV). The mirror assembly comprises four quadrants; each quadrant contains a $3\times3$ set of MPO plates and feeds one CMOS detector. The focal spot is defined by the intersection of the detector plane and the vector projecting from the center of curvature to the target (adapted from Figure 1 of \citealt{2022ZhangChenApJL}; rotated for presentation).
    ({\em Right} panel): Positional offsets between theoretical and measured PSF focal spot centroids. Red arrows indicate pre-calibration offsets; blue arrows show post-calibration results.}
    \label{fig:pointing_mpo}
\end{figure*}

\begin{figure*}
    \centering
    \includegraphics[width=\textwidth]{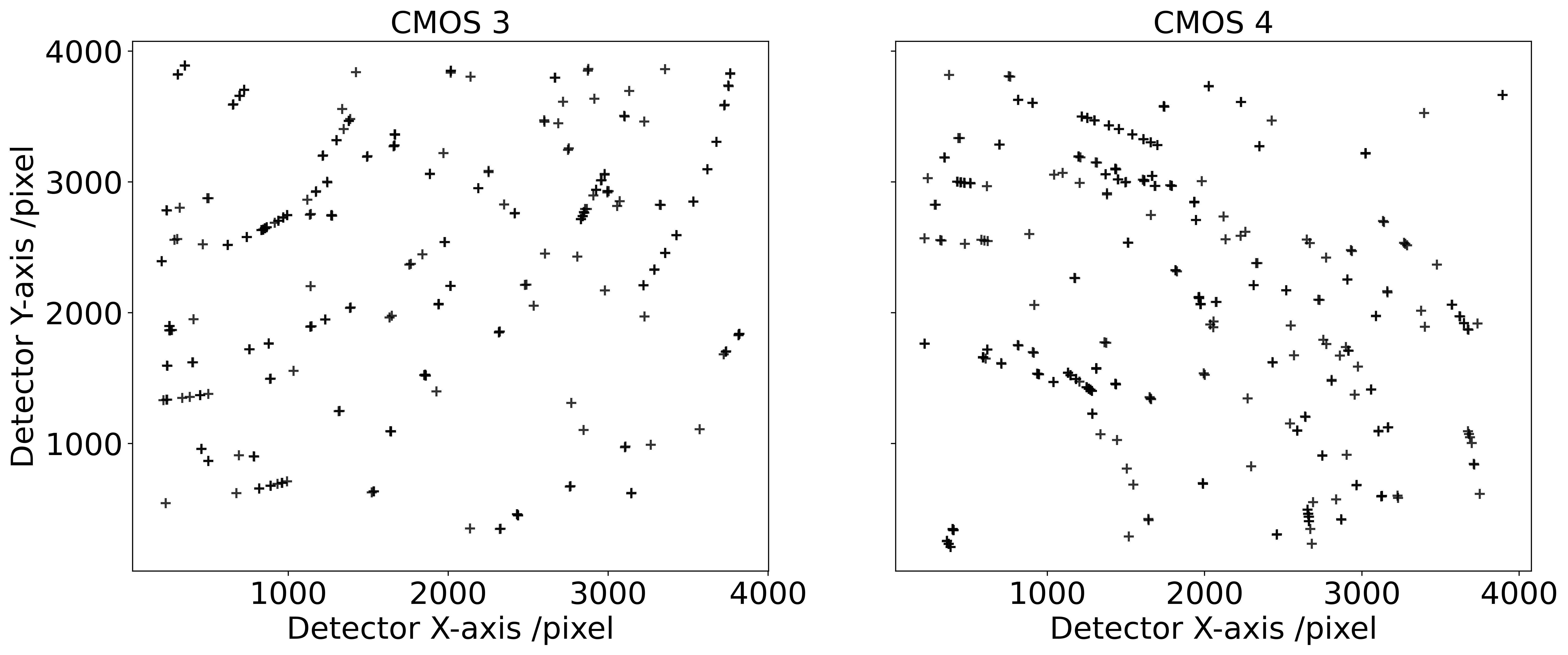}
    \caption{The positions of the source detections adopted for the characterization of the post-calibration source positional accuracy on the detector plane ({\em left} panel: CMOS 3, {\em right} panel: CMOS 4).}
    \label{fig:post_calibration_detection}
\end{figure*}

\begin{figure*}[!htbp]
    \centering
    \includegraphics[width=0.48\textwidth]{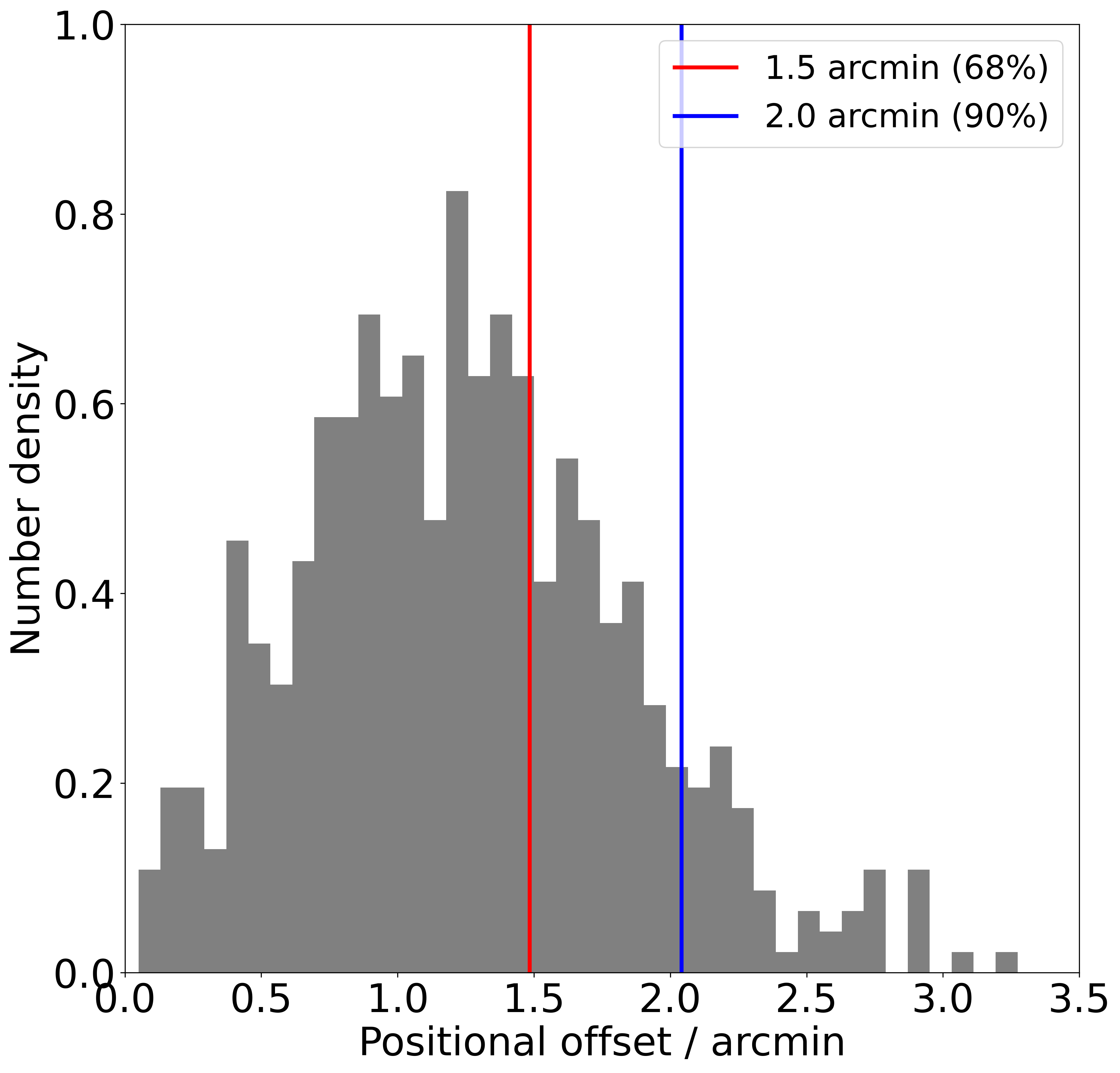}
    \includegraphics[width=0.47\textwidth]{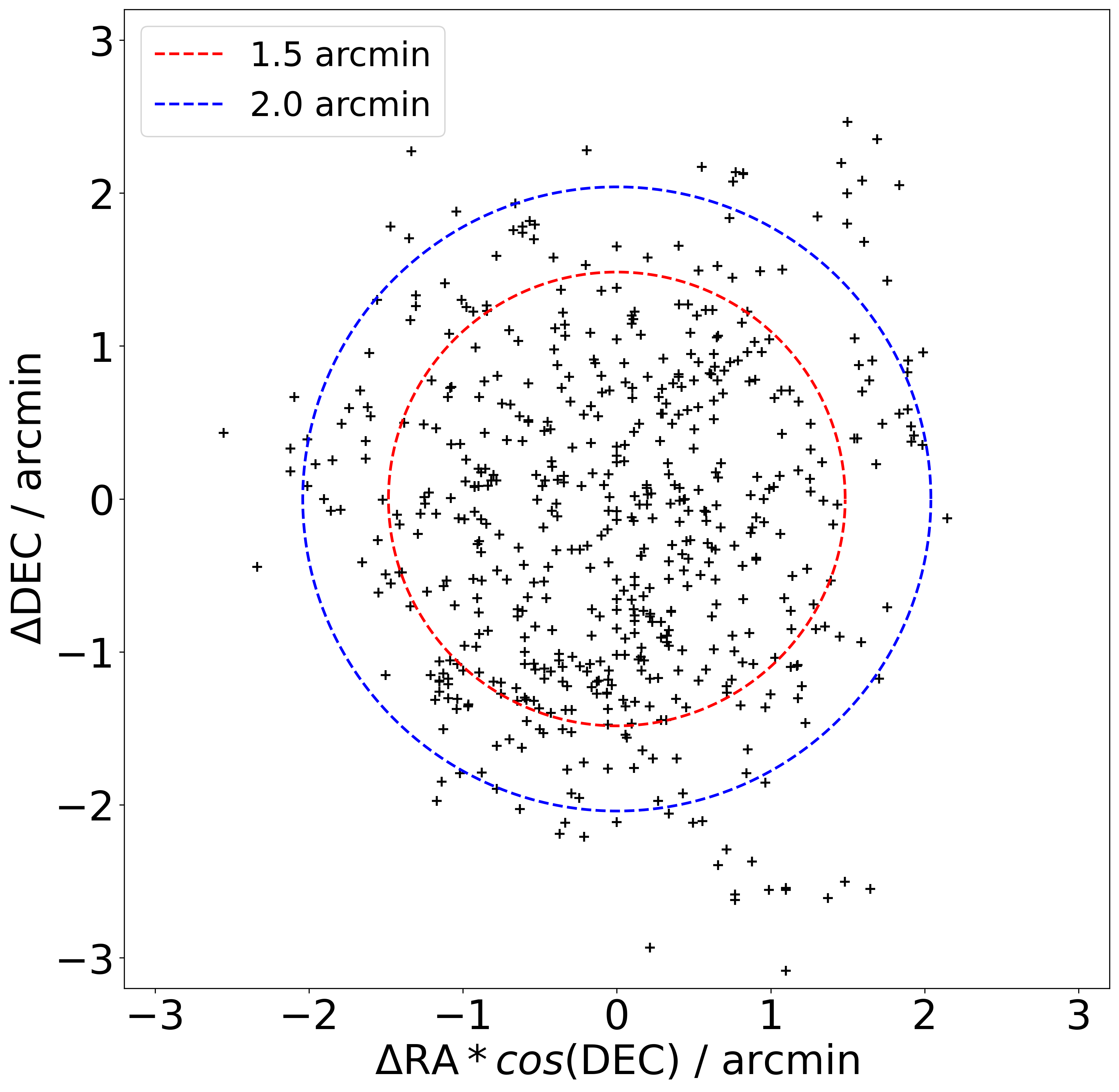}    
    \caption{({\em Left} panel) The histogram of the positional offset (units: arcmin) of the source detections adopted for the characterization of the post-calibration source positional accuracy. The 68 percentile ($1.5'$) and the 90 percentile ($2.0'$) are marked with red and blue solid lines, respectively. ({\em Right} panel) Positional errors displayed in two-dimensional form. The red and blue circles in dashed lines correspond to $1.5'$ and $2.0'$, respectively.}
    \label{fig:focalspot_offset}
\end{figure*}

Source positional calibration represents the highest-priority objective in \textit{LEIA}'s first in-orbit calibration phase. As the technological pathfinder for the \textit{EP} mission, \textit{LEIA}'s calibration will establish a robust paradigm for arcminute-level transient localization ($\leq2'$ at $90\%$ confidence level) across \textit{EP}-WXT's unprecedented FoV exceeding 3600 deg$^2$, the largest ever achieved by a focusing X-ray instrument. This pioneering methodology addresses the critical absence of established calibration standards for wide-field lobster-eye optics, providing essential reference data for next-generation X-ray missions.

Owing to constraints in experimental facilities and operational conditions, no pre-launch source positional calibrations were performed. Initial on-orbit observations during the Performance Verification (PV) phase revealed a relatively poor localization accuracy of $\geq10'$.
In general, the observed positional deviations arise primarily from three principal factors: 1) attitude determination uncertainties inherent to the single-star tracker configuration; 2) potential variations in the coordinate transformation matrices between the star tracker and \textit{LEIA} instrument, and those among different instrumental ingredients (MPO optics, detector platform and individual CMOS sensors), most likely induced by post-launch stress relaxation and thermal environmental changes; 3) randomly distributed residual offsets from imperfect PSF characteristics that form the non-linear correction matrix.
The calibration procedure addresses these issues through two complementary approaches: the refinement of the celestial-to-detector coordinate transformation matrix to correct systematic displacements (factors 1 and 2), and empirical modeling of residual non-linear distortions (factor 3) via the Radial Basis Function (RBF) interpolation algorithm \citep{Wahba1990Spline}. These corrections are incorporated into the updated telescope definition files of the CALDB.

In practice, the source positional calibration is performed by implementing a series of pointing observations of the Crab within the FoV, from October 2022 to February 2023.
For each FoV quadrant, a $11\times11$ sampling grid corresponding to 121 evenly distributed incident directions was adopted.
The main steps of the calibration procedure is briefly summarized as follows.
Firstly, with the Right ascension (R.A., $\alpha_{\rm c}$) and Declination (Dec., $\delta_{\rm c}$) in J2000 coordinates of the celestial object, we can derive the vector of the source direction $\overrightarrow{r_{J2000}}$ within the J2000 coordinate system,
\begin{equation*}
    \overrightarrow{r_{J2000}}=\left\{\begin{array}{c}
x_{J2000}=\cos \alpha_{\rm c} \cdot \cos \delta_{\rm c} \\
y_{J2000}=\sin \alpha_{\rm c} \cdot \cos \delta_{\rm c} \\
z_{J2000}=\sin \delta_{\rm c}
\end{array}\right.
\end{equation*}
This vector can then be expressed in the satellite coordinate system,
\begin{equation*}
    \overrightarrow{r_b}=R_{\rm SAT} \cdot \overrightarrow{r_{\rm J2000}}
\end{equation*}
Here $R_{\rm SAT}$ is the rotational matrix expressed by the quaternion of the satellite ($Q_0$, $Q_1$, $Q_2$ and $Q_3$) during the stable period of one pointing observation (the angular distance to the source $\rm ANG\_DIST<0.05\deg$).
Next, this vector is transformed to the detector coordinate system
by performing an additional transformation,
\begin{equation}
    \overrightarrow{r_{\text {DET}}}=M_{\rm SAT 2 DET} R_{\mathrm{C}} \overrightarrow{r_b}
\end{equation}
where $M_{\rm SAT 2 DET}$ is the transform matrix from the satellite coordinate system to the CMOS detector coordinate frame measured during the course of ground test and calibration.
The correction matrix $R_{\mathrm{C}}$, composed of three Euler angles $\alpha$, $\beta$ and $\gamma$ following an `X-Y-Z' rotation, is defined as 
$\begin{aligned} & R_{\mathrm{c}}=R_z(\gamma) \cdot R_y(\beta) \cdot R_x(\alpha) \\ & =\left(\begin{array}{cc}\cos \beta \cos \gamma & \sin \alpha \sin \beta \cos \gamma-\cos \alpha \sin \gamma \\ \cos \beta \sin \gamma & \sin \alpha \sin \beta \sin \gamma+\cos \alpha \cos \gamma \\ -\sin \beta & \sin \alpha \cos \beta\end{array}\right. \\ & \left.\begin{array}{c}\cos \alpha \sin \beta \cos \gamma+\sin \alpha \sin \gamma \\ \cos \alpha \sin \beta \sin \gamma-\sin \alpha \cos \gamma \\ \cos \alpha \cos \beta\end{array}\right)\end{aligned}$

The correction matrix $R_{\mathrm{C}}$ is set to an identity matrix (i.e. $\alpha=0,~\beta=0,~\gamma=0$) before the calibration.
The {\em theoretical} position of the focal spot ($X_{\rm theory}, Y_{\rm theory}$) is derived from lobster-eye optics principles by projecting $\overrightarrow{r_{\text {DET}}}$ onto the detector plane from the curvature center (Figure \ref{fig:pointing_mpo}, left panel).
On the other hand, through PSF centroid analysis we obtain the {\em measured} position of the focal spot ($X_{\rm measure}, Y_{\rm measure}$).
As an example, the quiver map of the positional offset within the FoV of CMOS 4 is presented in the right panel of Figure \ref{fig:pointing_mpo}.
Pre-calibration analysis (i.e. when the correction matrix is set to an identity matrix) revealed systematic positional offsets up to $\sim$78 pixels (equivalent to $\sim$$11'$) towards the negative X-axis direction, as denoted by the red arrows.
Least-square minimization ($\epsilon^2\equiv\Sigma_{i=1}^{121}\{(X_{i, \rm theory}-X_{i, \rm measure})^2+(Y_{i, \rm theory}-Y_{i, \rm measure})^2\}$) 
results in optimal Euler angles of [$-1.83\deg$, $-2.35\deg$, $7.73\deg$], reducing maximum displacements to $\sim$23 pixels ($\sim$$3.1'$), as denoted by the blue arrows.
In the meantime, the largely anisotropic distribution of residual offset is generally consistent with expectations for random pointing errors from inherent micro-pore channel distortions.
By employing the \texttt{scipy.interpolate.RBFInterpolator} tool\footnote{\url{https://docs.scipy.org/doc/scipy/reference/generated/scipy.interpolate.RBFInterpolator.html}}, the residual offset is modeled and stored as a $257\times257$ matrix in the CALDB.

Finally, the post-calibration positional accuracy of \textit{LEIA} is evaluated by comparing measured source positions with higher-accuracy reference positions from the SIMBAD astronomical database \citep{2000WengerSIMBAD}.
Data from CMOS 3 and CMOS 4 observations beginning in March 2023 are used, comprising an initial catalog of 6932 valid detections from 238 sources. To minimize potential biases that may dilute the conclusions, a few steps are taken to filter the parent sample.
Firstly, we discarded the detections from extended sources including the supernova remnants (SNRs), Cluster of Galaxies, Globular Cluster, Cluster of Stars and Open Cluster.
Secondly, the detections near the edge of the detectors (i.e. within a distance of 200 pixels) are also rejected, as the focal spot is prone to be distorted in these regions due to the vignetting effect.
Thirdly, we removed low-significance detections (defined as those with less than 100 counts in the focal spot region), to reduce statistical uncertainties.
Lastly, the four persistent X-ray binaries within the LMC region, including LMC X-1, LMC X-2, LMC X-3 and LMC X-4 are excluded from the sample. 
This is mainly due to the fact that the LMC region was regularly monitored by \textit{LEIA} on a daily basis \citep{2023LingZXRAA}. As a consequence, the positions of these sources on the detectors barely changes, introducing a non-negligible bias in the quantification of the averaged positional accuracy across the whole detector.
Our final catalog consists of 572 detections from 43 sources. The positions of these detections on the detector plane are shown in Figure \ref{fig:post_calibration_detection}, which exhibit a largely uniform distribution on both CMOS detectors. 
In Figure \ref{fig:focalspot_offset}, we present the distributions of the positional offset obtained from these detections.
The left panel shows the histogram of the positional deviations. 
It is found that, despite of a few cases where the positional deviation may reach as large as $3'$, the majority of the offset fall within $2.5'$ with a median value of $1.2'$.
The 68 percentile and 90 percentile are $1.5'$ and $2.0'$, respectively.
The two-dimensional positional offsets of the source detections are presented in the right panel of Figure \ref{fig:focalspot_offset}.
The systematic error of source localization is thus estimated to be $\sim2'$ at the 90\% confidence level, which meets the design requirement.
This confirms \textit{LEIA}'s capability in achieving its core scientific objectives.

\section{Effective area}
\label{sec:effarea}

\begin{figure*}[htbp]
    \centering
    \includegraphics[width=0.65\textwidth]{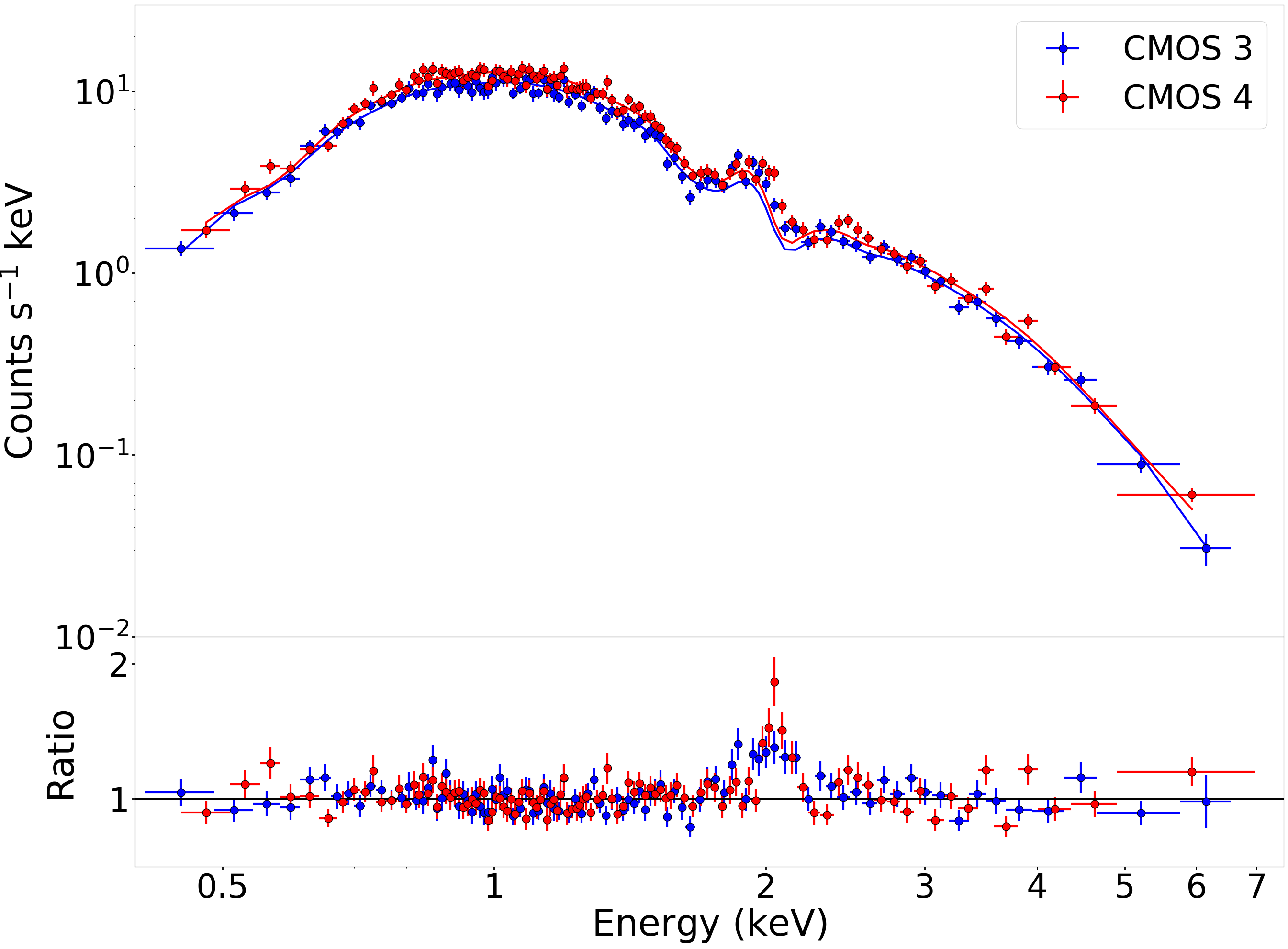}
    \caption{The Crab spectral analysis (0.4--8 keV) for the two observations carried out in the first calibration stage, with the source located on the centers of CMOS 3 (red symbols) and CMOS 4 (blue symbols).} \label{fig:crab_wxt15_wxt16_spectralfit_onaxis}
\end{figure*}

The Crab is a standard source for the calibration of the effective area for X-ray satellites \citep{Toor_Seward_1974, Kirsch2005, Weisskopf2010Crab, 2015Mason_NuSTAR_calibration}.
It has been acknowledged that within 1-100 keV band, the spatially phase-averaged integrated spectrum of the Crab nebula + centrally located pulsar can be very well described by a simple absorbed power law model with a column density of $\sim$$5\times10^{21}~{\rm cm^{-2}}$, a photon index of $\sim$2.1 and normalization of $\sim$10 \citep[for a detailed summary of the results obtained by different instruments, see][]{Kirsch2005}. 

We make use of the {\textsc{XSPEC}} software \citep{1996ASPC..101...17A} to perform the spectral analysis.
As a common practice, a simple absorbed power-law model (\texttt{tbabs*powerlaw}) is adopted to fit the time-averaged spectrum, where \texttt{tbabs} is the interstellar absorption using \textsc{WILMS} abundances \citep{2000ApJ...542..914W} and \textsc{BCMC} cross-sections \citep{1992ApJ...400..699B}. The differential photon spectral model is expressed by
\begin{equation}
    \frac{dN(E)}{dE}={E^{-N_{\rm H}\sigma(E)}}\times NE^{-\Gamma}~[{\rm ph~s^{-1}~{cm}^{-2}~{keV}^{-1}}]
\end{equation}
where $E$ is the photon energy, $\sigma(E)$ is the electron scattering cross section, $N_{\rm H}$ is the hydrogen column density, $\Gamma$ is the photon index and $N$ is the normalization factor. 

As an example, Figure \ref{fig:crab_wxt15_wxt16_spectralfit_onaxis} shows the data, best-fit model and fitting residuals of the 0.4--8 keV Crab spectra for CMOS 3 and 4 during the first stage of calibration, with the source imaging towards the center of the detectors. 
It is found that the overall spectra can be generally well reproduced by the absorbed power-law model, yielding statistically acceptable statistics of 
$\chi^2/{\rm d.o.f}=277/225$ and $296/232$, respectively.
A spectral residual is observed at $\sim2$ keV with a data to model ratio of $\sim1.5$--$2$, indicating an underestimation of the effective area at this energy. 
The Crab spectra of 242 individual observations on these two detectors are uniformly fitted with the aforementioned absorbed power-law model, yielding best-fit parameters of $N_{\rm H}=(5.23\pm0.81)\times10^{21}~{\rm cm^{-2}}$, $\Gamma=2.19\pm0.13$ and $N=11.36\pm2.18$.
The observed 0.5--4 keV flux is estimated to be $F_{\rm 0.5-4}=(1.95\pm0.23)\times10^{-8}~{\rm erg~s^{-1}~cm^{-2}}$. Note that the errors are all given at 90\% confidence level. These values agree largely with those reported by other X-ray satellites \citep[e.g.][]{1997ApJ...491..808P, 2000A&A...361..695M, Kirsch2005, 2022JATIS...8c4003M}, indicating that the in-orbit effective area is overall consistent with theoretical predictions, as well as ground measurements.

\begin{figure*}[htbp]
    \centering
    \includegraphics[width=0.65\textwidth]{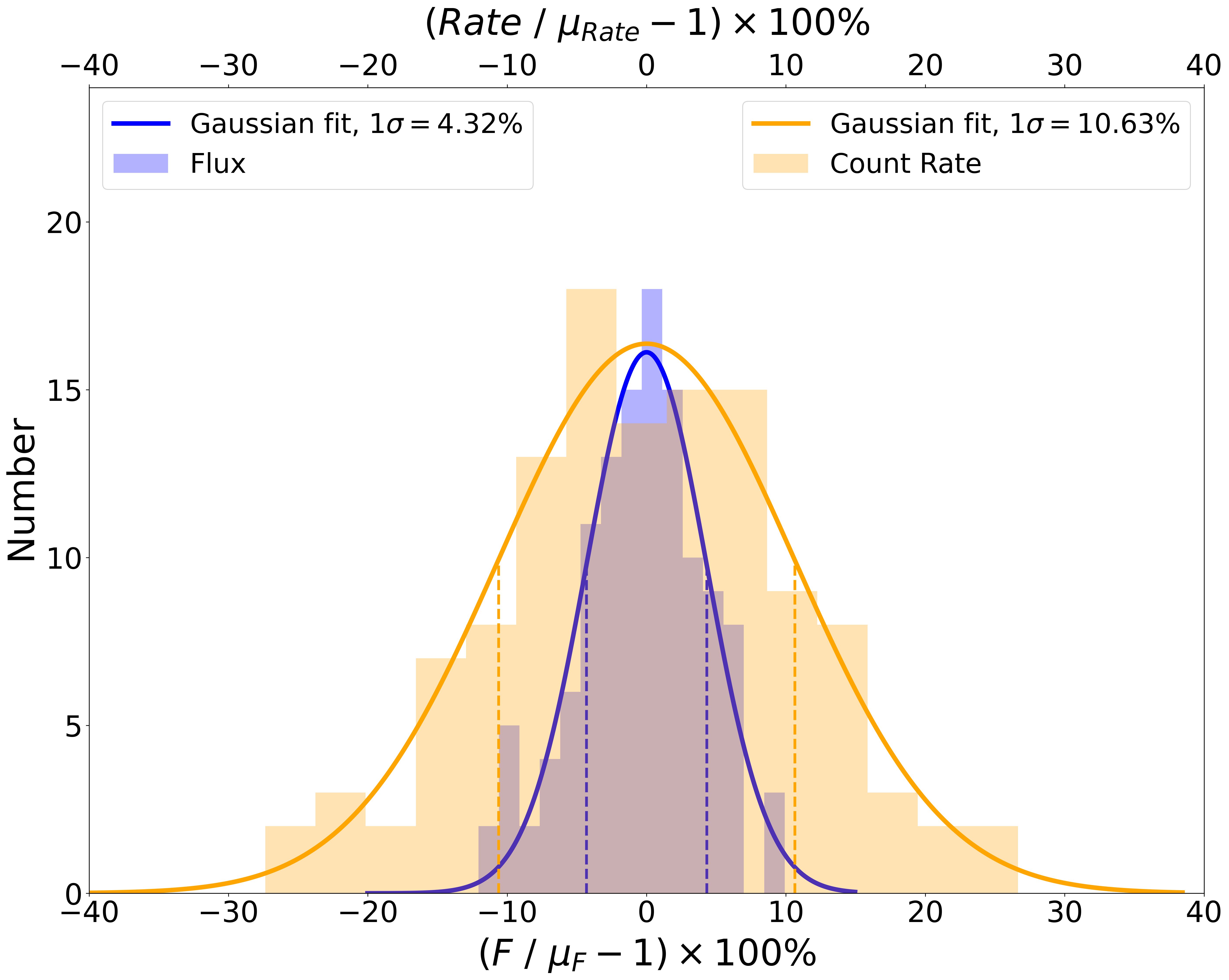}
    \caption{The distribution of the count rate (orange histogram) and observed flux (blue histogram) of the Crab nebula in 0.5--4 keV band, observed on CMOS 4, over-plotted with the best-fit Gaussian models (orange and blue lines). Both parameters are scaled by their mean values.}
    \label{fig:crab_cmos16_fluxrate}
\end{figure*}

We further evaluate the systematics of the effective area by combining the spectral fitting results obtained from different directions.
As an example, in Figure \ref{fig:crab_cmos16_fluxrate} we present the distributions of the observed count rate and flux (both within 0.5--4 keV band) derived from the best-fit absorbed power-law models for the $11\times11$ sampling directions of CMOS 4.
The Gaussian fits are performed for both distributions, yielding $1\sigma$ deviations of $\sim10.6\%$ for the count rate and $\sim4.3\%$ for the measured flux.
Specifically, the dispersion of the count rate distribution is a representative of the non-uniformity of the effective area within the FoV, which is mainly induced by several factors: the vignetting effect (predominant at the FoV edge due to the incompleteness of the optics beyond), the obscuration effect (in some directions the incident photons are blocked by the supporting structure of the MPO plates), and the inherent imperfections of the MPO devices introduced during the manufacturing process.
We note that this dispersion is largely consistent with that measured in ground calibration experiments (see Figure 9 of \citet{Cheng2024a}).
The dispersion of the measured flux, on the other hand, arises from the inherent discrepancy between the model and the real optical system, thereby signifies the systematics of the effective area.
For the FoV quadrant subtended by CMOS 3 the distribution of the flux is slightly broader, of $\sim8\%$. 
We thus conclude that the systematics for the effective area is $\lesssim10\%$ at $68\%$ confidence level.

\begin{figure*}
    \centering
    \includegraphics[width=0.65\linewidth]{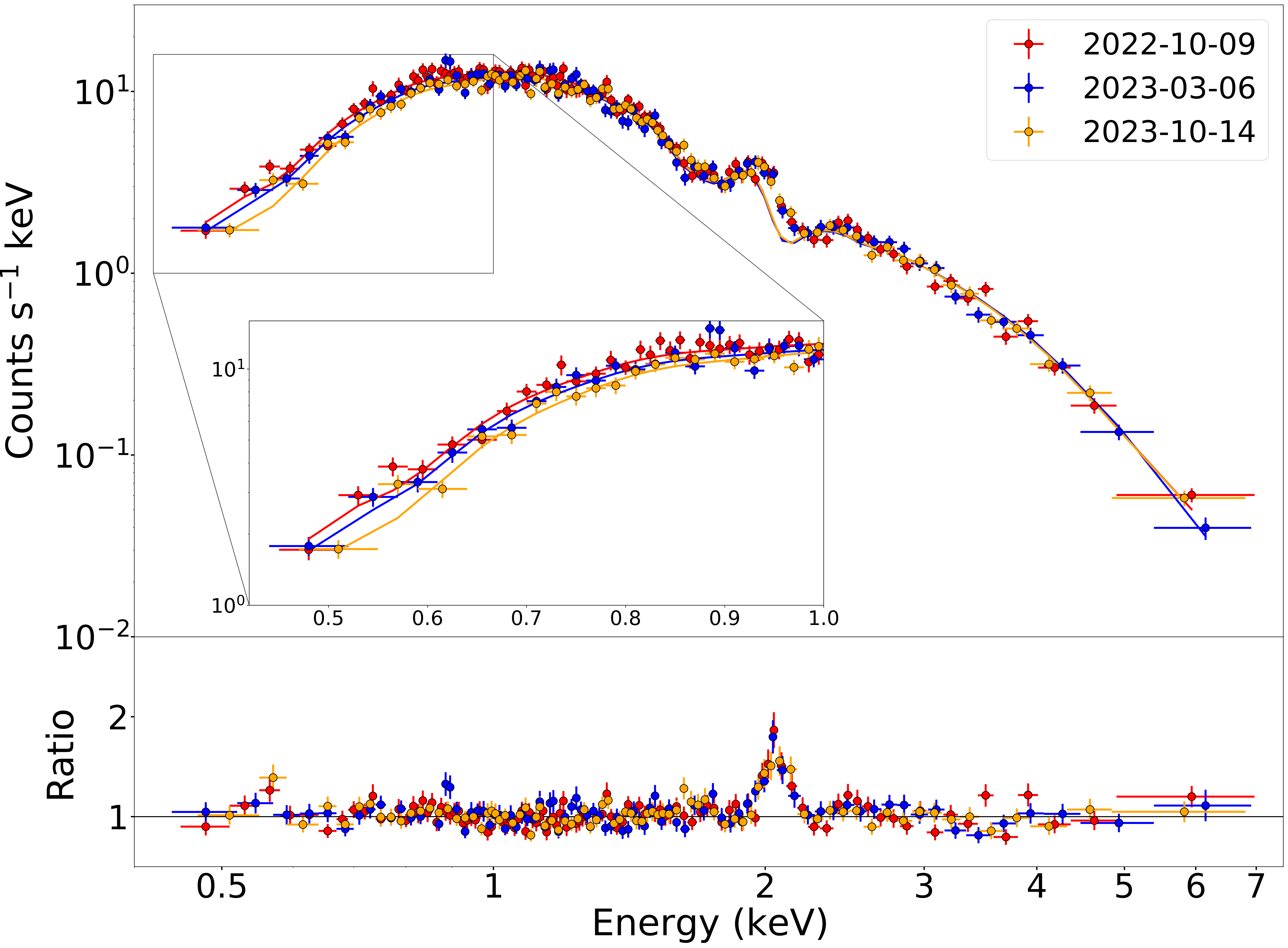}
    \caption{
    The Crab spectral analysis (0.4--8 keV) for the three observations carried out in October 2022 (red), March 2023 (blue) and October 2023 (orange), respectively. The source is located on the center of CMOS 4 in these observations. The spectra at 0.4--1.0 keV is zoomed in for a better illustration.}
    \label{fig:effarea_longterm}
\end{figure*}

The continuous monitoring of the effective area variation is a critical topic of X-ray missions, as it is found that the accumulation of contaminants on optical elements and detector surfaces could significantly deteriorate low-energy performance \citep{O_Dell2010}.
This has been a long-standing issue for the {\it Chandra} X-ray Observatory since its launch.
Organic compounds, likely originating from volatile materials within the spacecraft, condense on the Advanced CCD Imaging Spectrometer (ACIS) optical blocking filters, forming a layer that absorbs low-energy X-rays. 
This degradation has prompted routine calibration efforts and contamination modeling to mitigate its impact \citep[e.g.][]{Marshall2004, Plucinsky2022, Grant2024}. 
A moderate contamination issue was also noticed in the {\it XMM-Newton} mission \citep[e.g.][]{Kirsch_2005}.
We investigate the potential effective area deterioration, by conducting three consecutive observations of the Crab at different calibration stages.
In Figure \ref{fig:effarea_longterm}, we show the fittings of the Crab spectra obtained in three separate calibration stages. The source was located at the center of CMOS 4 in these observations, which were carried out in October 2022, March 2023 and October 2023, respectively.
Although the spectral fitting results remain largely consistent, we observe a continuous decline in the source count rate at low energies (0.4--0.8 keV) across the three observational epochs. 
Specifically, the measured count rate in October 2023 exhibits a $\sim15\%$ reduction compared to the initial October 2022 observation, 
accompanied by an increase in the best-fit hydrogen column density from $5.6_{-0.2}^{+0.2}\times10^{21}~{\rm cm^{-2}}$ to $6.2_{-0.2}^{+0.2}\times10^{21}~{\rm cm^{-2}}$.
This evolutionary trend indicates progressive contaminant accumulation that deteriorates the low-energy effective area.
Extrapolating this contamination rate under equivalent environmental controls would predict a $\sim$$50\%$--$80\%$ degradation in the effective area below 1 keV for \textit{EP}-WXT, over its nominal 3–5 year operational period.
To track this, low-energy performance monitoring will be implemented as a routine operational procedure.

While the observed spectra of the Crab can be overall reproduced by the absorbed power-law model with parameters generally in agreement with those reported in the literature, there is a clear spectral residual around $2$ keV (see Figure \ref{fig:crab_wxt15_wxt16_spectralfit_onaxis}), indicating an underestimation of the effective area at this energy.
We note that this mismatch between the data and model is also found in the ground calibration experiment of the mirror assembly conducted at MPE/Panter X-ray test facility \citep{Rukdee2023}. 
Though the nature of this narrow-band residual remains poorly understood yet, we suggest it is most likely to be associated with the imperfect modeling of the effective area near the energy of iridium (Ir) absorption edge \citep{Gorenstein2010}.
For both {\em LEIA} and \textit{EP}-WXT instruments, the micro pores of the mirror assemblies were coated with iridium to enhance X-ray reflectivity \citep{2023OptMa.14214120L}. 
Theoretically, for pure iridium there is a cascade of reflectivity decrement due to the iridium absorption around 2 keV (Ir-M5, 2.04 keV; Ir-M4, 2.12 keV), which results in a sharp decrease of the effective area around this energy (from $\sim1.95~{\rm cm^{2}}$ at $\sim1.9~{\rm keV}$ to $\sim0.1~{\rm cm^{2}}$ at $\sim2.05~{\rm keV}$).
The spectral fitting results can be understood in the sense that the actual
effective area exhibits a more gentle decrease across the absorption edge.
This may be due to the fact that the purity of iridium is not as high as theoretically predicted, suggesting potential presence of impurities or doped elements within the Ir material.
In fact, previous studies suggest that with the over-layer of a complementary reflective material, e.g. Chromium (Cr), the reflectivity reduction at 2 keV can be compromised \citep{Stehlikova2019, Dohring2021}.
In this sense, it is promising to mitigate the spectral residual by involving additional coating elements/layers when modeling the effective area curve.
This is now in progress and will be presented in a separate paper (Zhao et al. in preparation).

\section{Energy scale and spectral resolution}
\label{sec:cas_a_data_analysis}

\begin{figure*}[!htbp]
    \centering
    \includegraphics[width=0.47\textwidth]{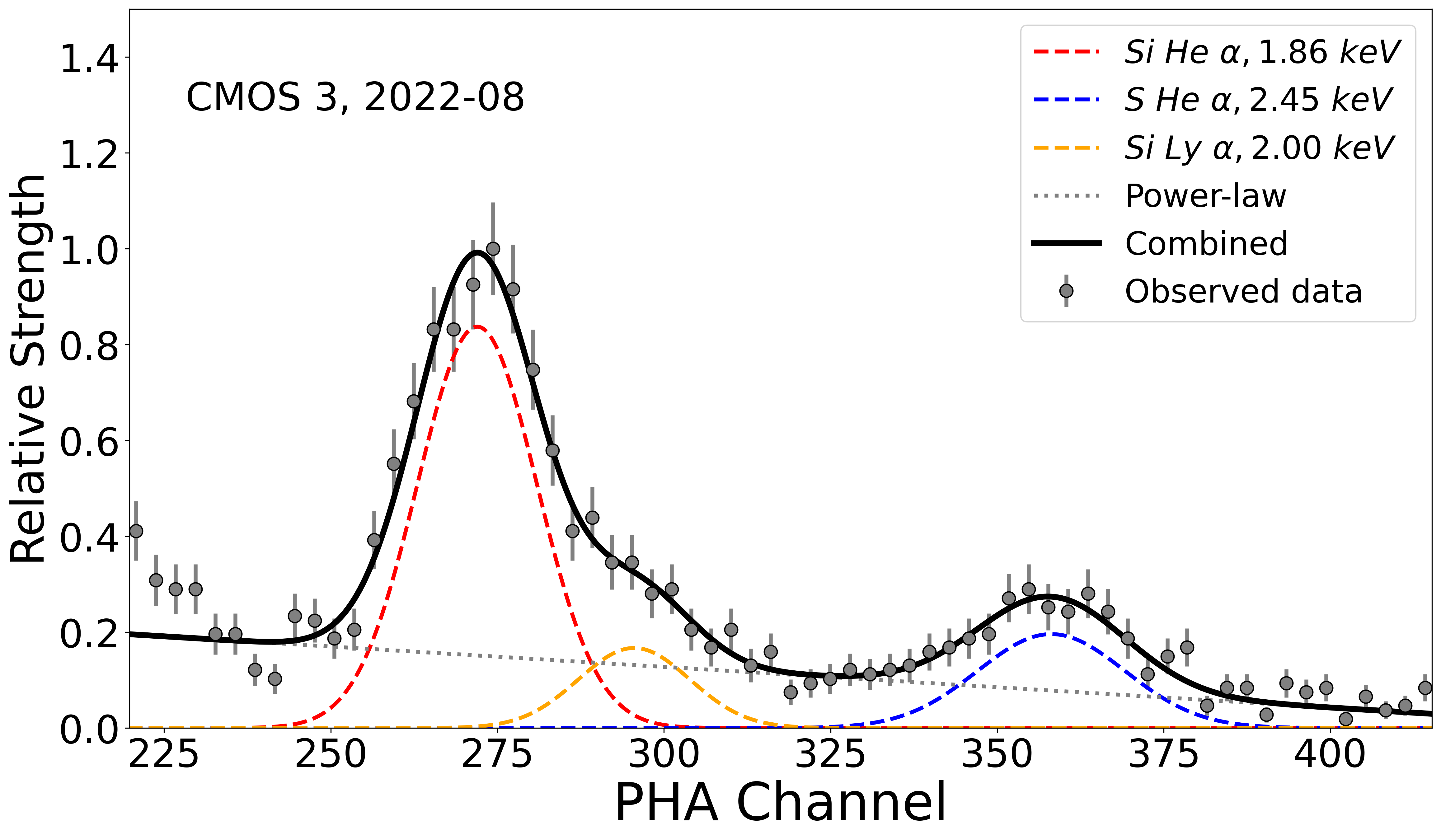}
    \includegraphics[width=0.48\textwidth]{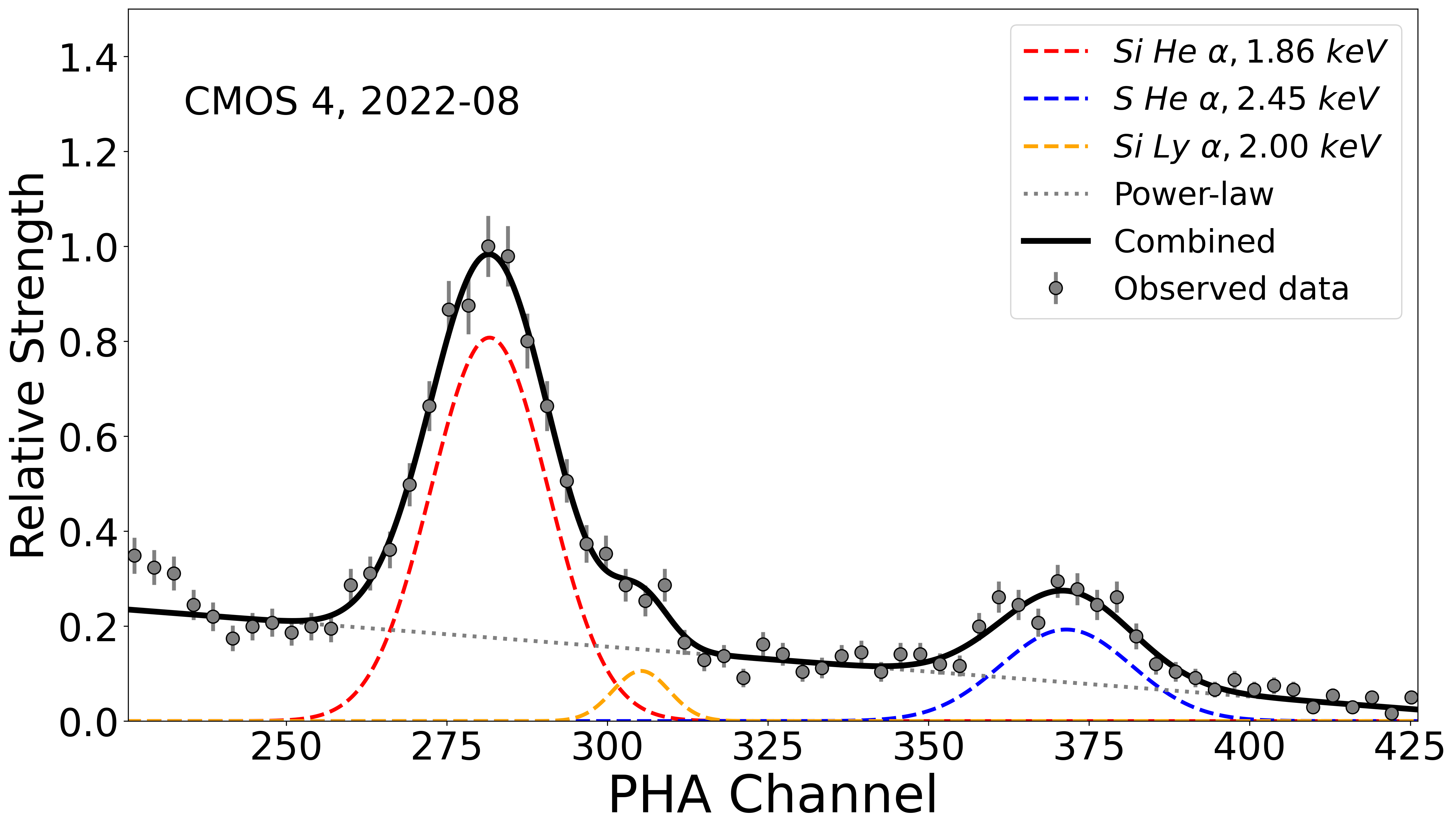}
    \caption{Spectral analysis of the Cas A for the observations carried out in August 2022, with the source located at the center of CMOS 3 ({\em Left} panel) and CMOS 4 ({\em Right} panel). In each panel the data is denoted by the grey symbols, the underlying power-law continuum is signified by the gray dotted line, and the three emission lines (Si He-$\alpha$, S He-$\alpha$ and Si Ly-$\alpha$) are denoted by the red, blue and orange dashed lines, respectively. The combination of these components is donoted by the black solid line.}
    \label{fig:casa_spectralfit_aug2022_cmos16}
\end{figure*}

\begin{figure*}[!htbp]
    \centering
    \includegraphics[width=0.49\textwidth]{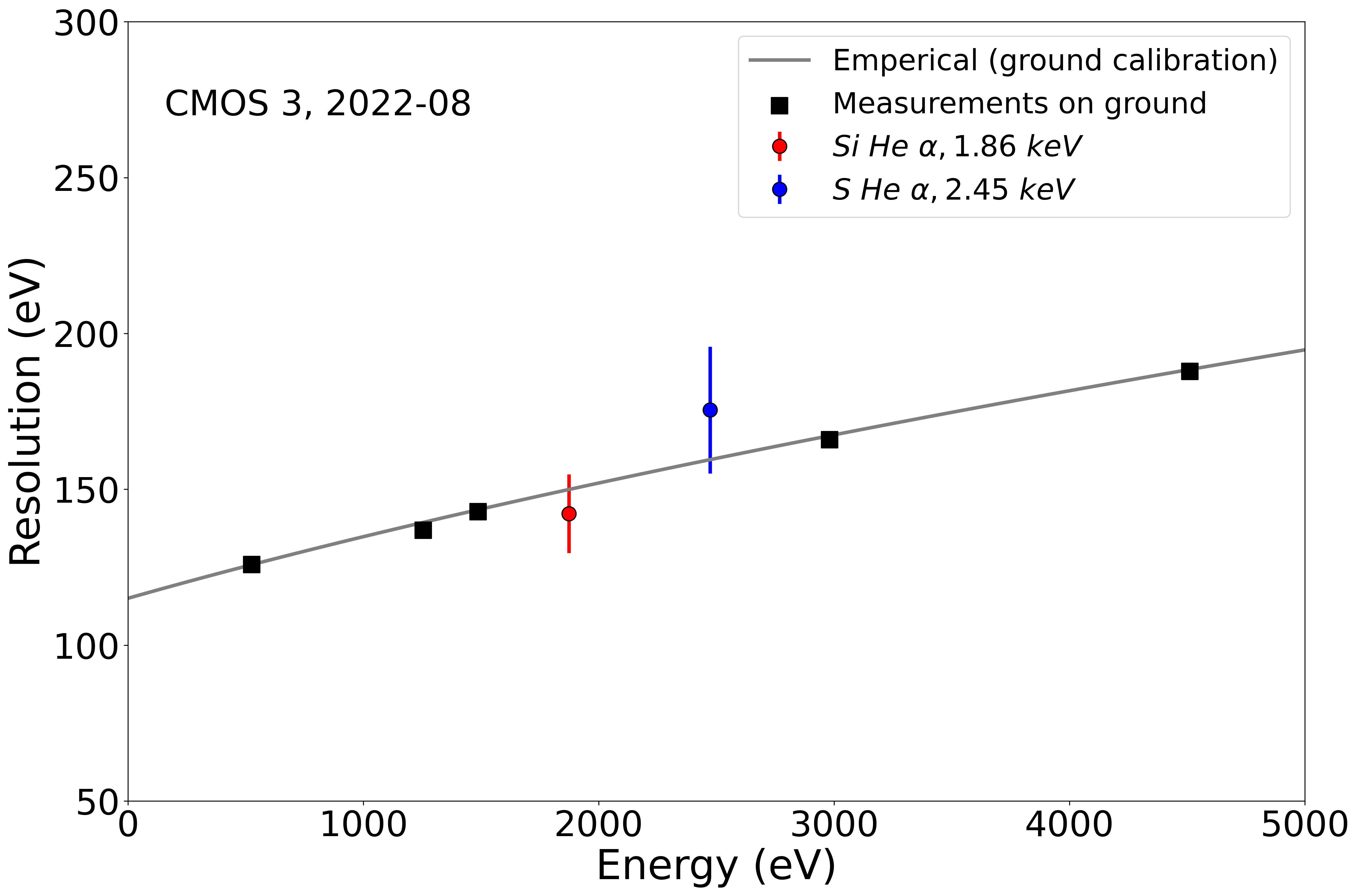}
    \includegraphics[width=0.49\textwidth]{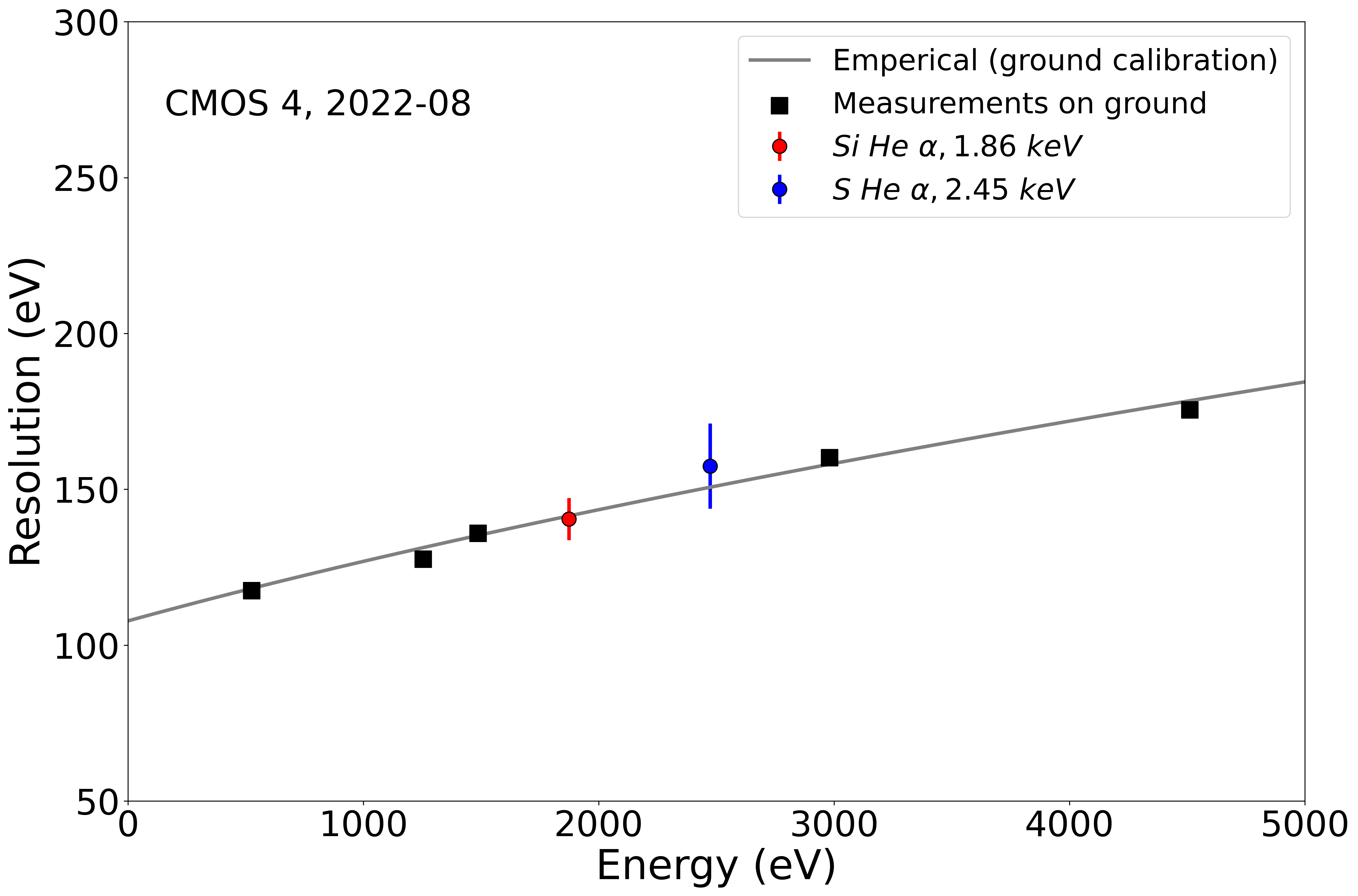}
    \caption{
    The relation between the photon energy and the spectral resolution, for CMOS 3 ({\em Left} panel) and CMOS 4 ({\em Right} panel). In each panel, the in-orbit spectral resolution at the energies of Si He-$\alpha$ and S He-$\alpha$, measured utilizing the data of August 2022, are denoted by red and blue dots, respectively. 
    The resolution measured from on-ground calibration experiments at different energies of characteristic X-ray emission lines (including O K$\alpha$, Mg K$\alpha$, Si K$\alpha$, Ag L and Ti K$\alpha$) are denoted by black squares. The empirical relation obtained from fitting Equation \ref{eq:fwhm_e} to the ground measurements is denoted by the black solid line.}
    \label{fig:casa_enresolution_aug2022_cmos16}
\end{figure*}

Before the launch of \textit{LEIA}, the energy response of the CMOS detectors was calibrated in details at the 100-m X-ray facility at IHEP using a multi-target electron impact X-ray source \citep{2023WangYusa}. The X-ray emission lines used for the establishment of the EC relation include O K$\alpha$ ($0.53$ keV), Mg K$\alpha$ ($1.25$ keV), Si K$\alpha$ ($1.74$ keV), Ag L$\alpha$ ($2.98$ keV), Ti K$\alpha$ ($4.51$ keV) and Ti K$\beta$ ($4.93$ keV). 
A good linearity between the energy and PHA channel was found for all four sensors onboard \textit{LEIA}, with slight differences in the slopes and intercepts. The spectral resolutions are $\sim120-140$ eV at 1.25 keV (Mg K$\alpha$) and $170-190$ eV at 4.51 keV (Ti K$\alpha$), meeting design requirements. 
Also, the gain coefficient and spectral resolution both exhibit a good spatial uniformity for each sensor \citep[see Section 3.3 in][for more details]{Cheng2024a}.
During the in-orbit calibration campaign, Cas A is used to characterize the potential variations in the energy response properties. 
The first Cas A observation was carried out in August 2022, a few weeks after the launch.
Later in February 2023 and October 2023, Cas A was observed twice to investigate the evolutionary trend of the detector performance after half-year and one-year operations. 
All these observations were carried out on CMOS 3 and CMOS 4, with the source located on the center of each detector.

For each calibration stage, all available spectra are stacked to increase the signal-to-noise ratio. 
The adjacent PI channels are also binned for achieving higher statistics.
Figure \ref{fig:casa_spectralfit_aug2022_cmos16} shows the observed 1.5--2.8 keV spectra of Cas A obtained in 2022 August on CMOS 3 ({\em Left} panel) and CMOS 4 ({\em Right} panel).
Three prominent emission lines, including Si He-$\alpha$ ($1.86$ keV), Si Ly-$\alpha$ ($2.0$ keV) and S He-$\alpha$ ($2.45$ keV), are clearly detected in the spectrum.
Note that the weaker Si Ly-$\beta$ emission line ($2.2$ keV) cannot be reliably detected in the data due to sensitivity limitations.
The spectral fitting is performed by employing a model composed of four components, including a power-law underlying continuum and three Gaussian components. 
This provides a statistically acceptable fit to the data ($\chi^2/{\rm d.o.f}=61/53$ and $64/53$, respectively), as shown in Figure \ref{fig:casa_spectralfit_aug2022_cmos16}.

The center channels and the widths of the three Gaussian components are obtained.
The in-orbit gain coefficient is evaluated by establishing the linear relation between the center channels and line energies, i.e. $E=k_{\rm gain}\times C+b$. 
For the Epoch-1 observation of August 2022, the gain values ($k_{\rm gain}$) are found to be $6.85\pm0.09$ and $6.57\pm0.10$ for CMOS 3 and CMOS 4, respectively, which are broadly consistent with ground values ($6.70\pm0.03$ and $6.54\pm0.04$, respectively).
On the other hand, the spectral resolutions at the energies of two stronger emission lines, Si He-$\alpha$ and S He-$\alpha$ are derived from the FWHMs of the corresponding Gaussian components\footnote{We note that the width of the Si Ly-$\alpha$ cannot be well constrained due to an insufficient S/N ratio, thereby is not included in the following analysis.}
(note that the intrinsic widths of these two emission lines can be neglected), with values of $142\pm13$ eV and $175\pm20$ eV for CMOS 3 and $140\pm7$ eV and $157\pm14$ eV for CMOS 4.

In Figure \ref{fig:casa_enresolution_aug2022_cmos16}, we present the in-orbit spectral resolutions measured at the energies of Si He-$\alpha$ and S He-$\alpha$ for CMOS 3 and CMOS 4 (denoted by colored dots), along with the ground calibration results (denoted by black squares) obtained at 100-m X-ray facility.
For the ground measurements, we formulate the relation between the photon energy ($E$, in units of eV) and energy resolution ($FWHM_{\rm E}$, in units of eV) with an empirical function \citep{Holland2013},
\begin{equation}
    FWHM_{\rm E}=2.35\omega(\sigma^2+f_{\rm Fano}E/\omega)^{1/2}
    \label{eq:fwhm_e}
\end{equation}
where $\omega$ signifies the average ionization energy and is set to $3.65$ eV, $f_{\rm Fano}$ represents the Fano factor \citep{Fano1947} and $\sigma$ represents the equivalent noise charge.
The best-fit relations, with optimal values of $f_{\rm Fano}\approx0.2$ and $\sigma\approx12$, are also plotted in Figure \ref{fig:casa_enresolution_aug2022_cmos16}.
It is found that the in-orbit spectral resolution is generally consistent with that predicted by on-ground measurements.

To assess the long-term variation of the energy response, we analyze the Cas A spectra of the Epoch-2 and Epoch-3 observations carried out in 2023 in a similar way.
No significant variations are observed for both the gain coefficient and spectral resolution, as expected.
We thus conclude that the energy response of the two detectors (CMOS 3 and CMOS 4) remains stable after one year of operation.
The results of the gain coefficient obtained from all three epochs are summarized in Table \ref{table:cmos_specifications}.

\begin{table*}[hbtp]
\caption{The gain coefficient 
of CMOS 3 and CMOS 4, measured at different stages of calibration campaign.}  
\label{table:cmos_specifications}      
\centering                          
\begin{tabular}{cccccc}        
\hline                
CMOS ID & 2021-11 (ground) & 2022-08 & 2023-02 & 2023-10 \\ 
\hline
3  & $6.70\pm0.03$ & $6.85\pm0.09$ & $6.83\pm0.04$ & $6.83\pm0.01$ \\
4 & $6.54\pm0.04$ & $6.57\pm0.10$ & $6.70\pm0.34$ & $6.49\pm0.18$ \\
\hline
\end{tabular}
\end{table*}

\begin{figure*}[!htbp]
    \centering
    \includegraphics[width=0.8\textwidth]{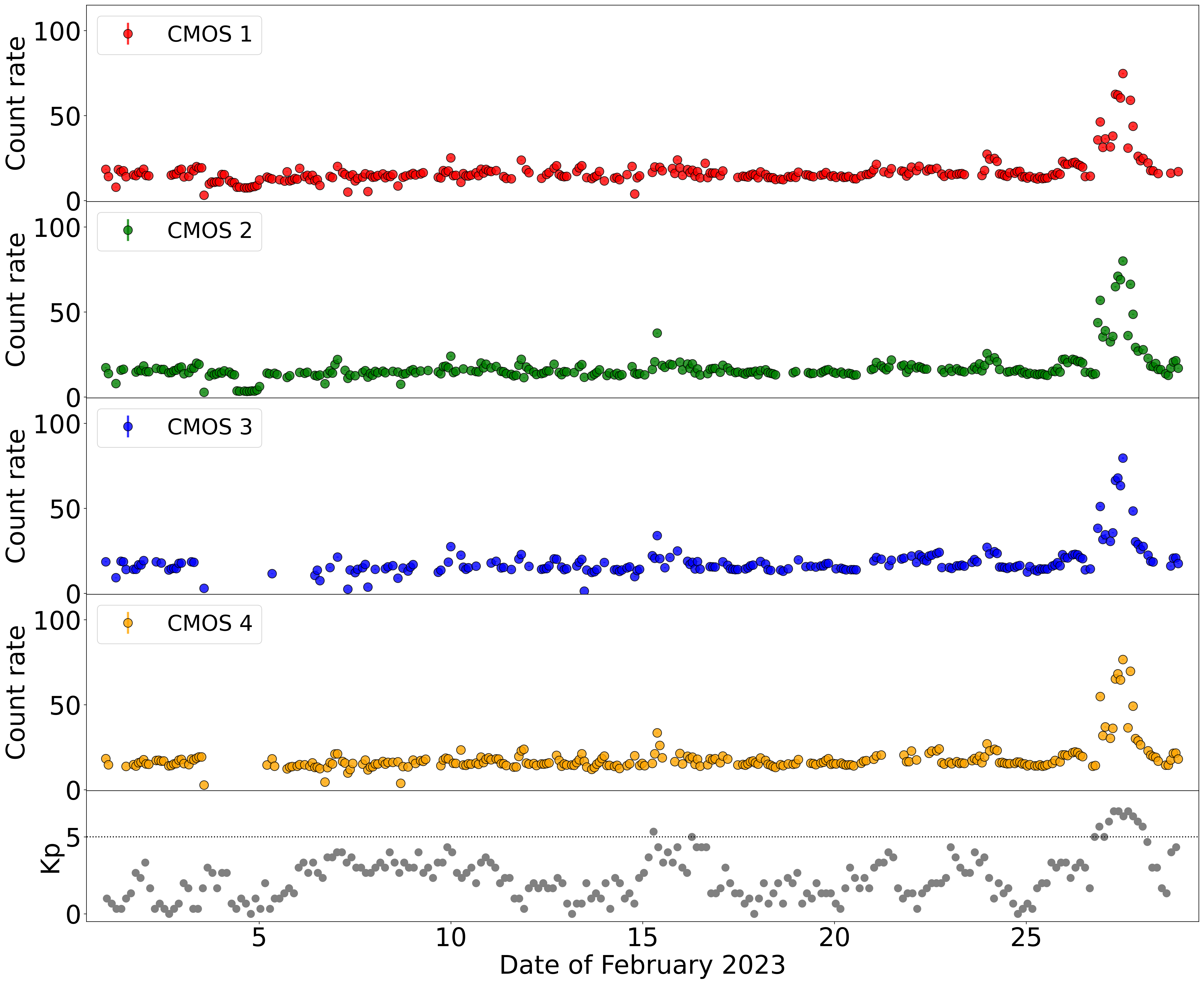}
    \caption{From upper to bottom: the 0.4--8 keV background count rate of CMOS 1, CMOS 2, CMOS 3 and CMOS 4, and the \kp~ parameter, as a function of time during February 2023. The dotted line in the lowest panel denotes the threshold of strong geomagnetic activity (\kp =5).}
    \label{fig:background_evolution_2023feb}
\end{figure*}

\begin{figure*}[!htbp]
    \centering
    \includegraphics[width=0.75\textwidth]{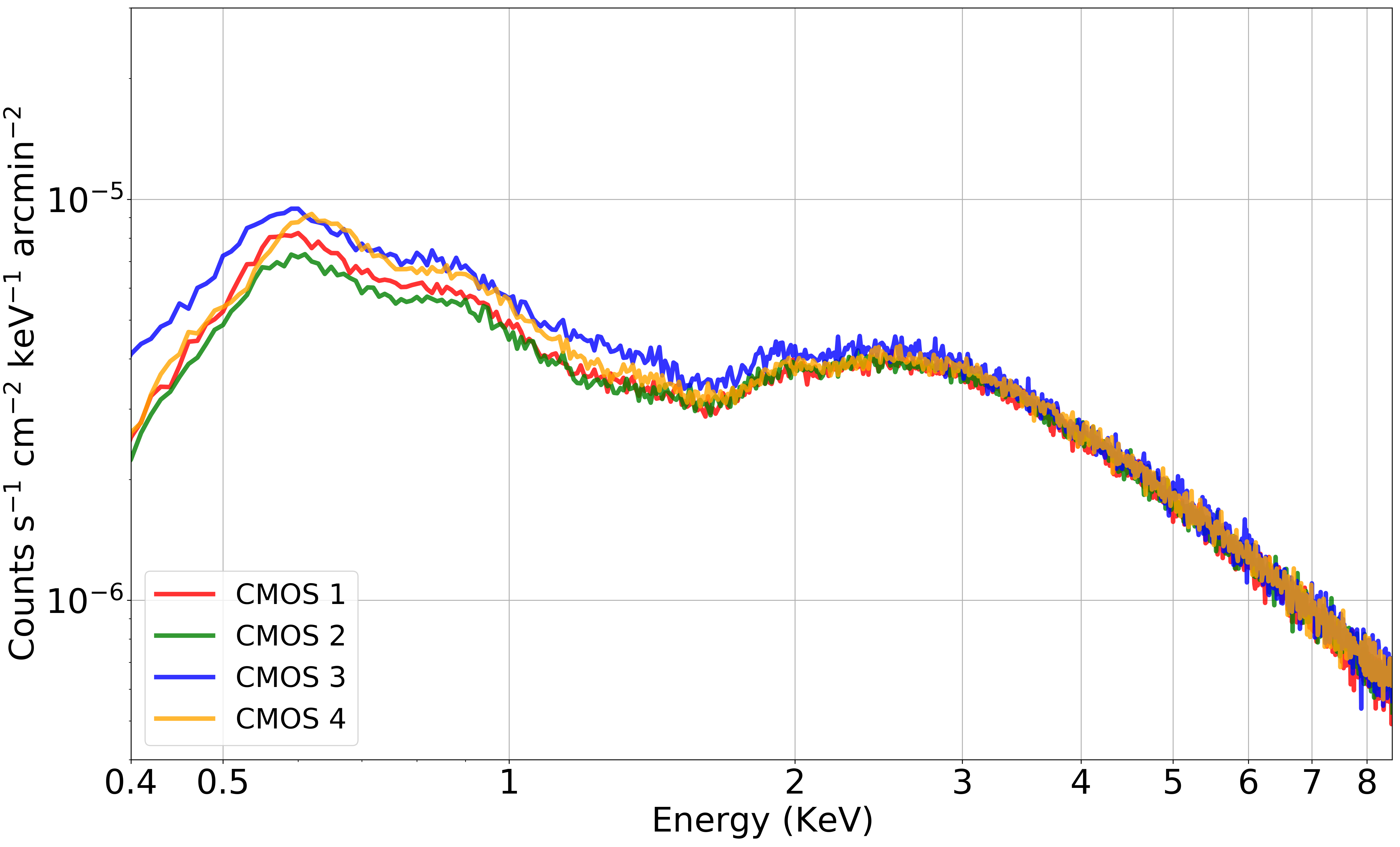}
    \caption{The time-averaged background count rate spectra of the four detectors onboard {\em LEIA} with observations conducted between February 1 and February 12, 2023 (no strong geomagnetic activities were present during this period).}
    \label{fig:leia_background_spectrum}
\end{figure*}

\section{Background}
\label{sec:background}

The satellite of {\em LEIA} is operating in a Sun-synchronous orbit of 500 km height, with an orbital period of 95 minutes and inclination angle of 98 degrees \citep{2023LingZXRAA}. 
Observations were conducted during the Earth's shadow for protecting the instrument and minimizing the impact of the solar radiation (the solar avoidance angle is set to 90 degrees). 
When the satellite passed through the South Atlantic Anomaly (SAA) region and south and north polar regions of the Earth, the background of the detectors is extremely high.
These time intervals are excluded from the data analysis.

Geomagnetic activity, conventionally characterized by the planetary K-index \citep[\kp,][]{Bartels1939Kp, Matzka2021KpReview}, plays a crucial role in modulating the background levels of on-orbit X-ray telescopes. During periods of strong geomagnetic activities (\kp $\ge$5), enhanced fluxes of charged particles, such as electrons and protons, from the Earth's magnetosphere or solar activity (e.g., solar energetic particles) can enter upper terrestrial atmosphere and satellite orbits. 
These particles may interact with X-ray detectors, leading to elevated particle-induced backgrounds.
This modulation effect is particularly significant during geomagnetic storms and at high geomagnetic latitudes, where satellites may encounter stronger particle flux due to Earth's magnetospheric configuration \citep[e.g.][]{Dudnik2023}.

In Figure \ref{fig:background_evolution_2023feb}, we present the evolution of the background count rate (in units of counts s$^{-1}$) within 0.4--8 keV of the whole detector region for CMOS 1--4 during February, 2023. 
The data are taken from observations during which the satellite was in low geographic latitude ($-15\deg$ < LAT < $15\deg$),
and those covering essentially blank sky regions without bright X-ray sources.
The latter is verified by cross-matching with the \textit{LEIA} all-sky survey catalogue (Hu et al. in preparation) and visually inspecting the observational images.
The evolution of \kp, taken from the National Oceanic and Atmospheric Administration website\footnote{\url{https://www.swpc.noaa.gov/products/planetary-k-index}} is plotted for comparison. It is found that the detector background is well correlated with the strength of the geomagnetic activity. Specifically, during the stable stage (\kp~$<5$), the count rate of the background events remains fairly stable, ranging from 10--20 counts/s. Accompanied by a strong geomagnetic activity between February 27 and 28 (\kp~$\ge5$), there is a prominent enhancement in the instrumental background exhibiting a count rate of $\sim$70--80 counts/s. 
In the meantime, the four CMOS detectors show similar behaviors, including the mean background count rate and the overall evolving trend. 
This indicates largely consistent performances of these four detectors within the first year of operation.

In Figure \ref{fig:leia_background_spectrum}, we present the averaged 0.4--9 keV background count rate spectra of the four CMOS detectors onboard {\em LEIA}, with the data taken from February 1 to 12, 2023, i.e. during when there were no strong geomagnetic activities.
Note that the spectrum is scaled by the nominal effective area of $3~{\rm cm^2}$ \citep{Cheng2024a}.
It is found that the background spectra exhibit a double-peak profile from low to high energies, which are qualitatively consistent with the simulated spectrum obtained in our previous work \citep{2018ZDH_WXTbackground} using the Geant4 toolkit \citep{2003AgostinelliGeant4}.
The spectra of the four CMOS sensors are overall consistent, particularly at the higher energy band where energetic particles dominate the emission \citep{2018ZDH_WXTbackground}. 
On the other hand, the portion below $\sim$1.5 keV exhibits slight differences, attributed mainly to the distinct sky coverages among different detectors.
We note that a more comprehensive investigation on the background properties of the four CMOS sensors aboard \textit{LEIA} is beyond the scope of this paper and will be presented elsewhere (Zhao et al. in preparation).

\section{Summary and Conclusion}
\label{sec:summary}

In this paper, we present comprehensive results from the in-flight calibration campaign of \textit{LEIA}, a pathfinder for the \textit{EP} satellite. During its two and half years of operations, we conducted a series of calibration observations across three distinct stages to characterize key instrumental properties, including the PSF, source localization accuracy, effective area and detector energy response (gain and spectral resolution). Primary calibration sources included Sco X-1 (for PSF), the Crab (for PSF, effective area and source localization) and Cas A (for energy response). 
The initial and most extensive calibration phase spanning from August 2022 to February 2023 delivered a thorough and comprehensive characterization of instrumental performance, successfully completing the highest-priority source localization calibration.
The second and third rounds of calibrations implemented in 2023 focused on long-term stability monitoring with reduced observational time requirements. 
Instrumental background properties were additionally investigated using scientific data acquired in February 2023.
The main findings from these calibration efforts are:

1) The imaging quality revealed by PSF observations of Sco X-1 and Crab exhibits no post-launch degradation, with PSF size and profiles consistent with ground measurements. The spatial resolution (represented by the FWHM along the long axis of the focal spot) ranges from $3.6'$--$9.3'$ with a median of $5.9'$.

2) The post-calibration source localization accuracy achieves $2'$ (J2000; 90\% C.L.) through the refinement of the celestial-to-detector coordinate transformation matrices and modeling of residual nonlinear offsets, which meets the design requirement.

3) The Crab spectra are well fitted by canonical absorbed power-law models with parameters of 
$N_{\rm H}=(5.23\pm0.81)\times10^{21}~{\rm cm^{-2}}$, $\Gamma=2.19\pm0.13$ and $N=11.36\pm2.18$, consistent with literature values.
The in-orbit effective area thus agrees with model predictions and ground measurements, though the residuals near 2 keV indicate required refinements at iridium absorption edges. The systematics at $68\%$ C.L. is estimated to be $\lesssim10\%$. Progressive contamination caused a $\sim15\%$ deterioration in low-energy effective area after one year of operation.

4) Cas A spectral analysis demonstrates that the in-orbit energy scale and spectral resolution are consistent with on-ground values. These parameters do not exhibit noticeable variations after the one year of operation.

5) An analysis of the instrumental background utilizing the data acquired in February 2023 reveals similar count rate evolutionary behavior across the various CMOS detectors. Moreover, the data exhibit robust geomagnetic modulation effects and demonstrate a qualitative spectral consistency with our previous simulations. 

As the pioneering operational lobster-eye X-ray telescope with a considerably large FoV ($18.6\deg\times18.6\deg$), \textit{LEIA} has successfully established rigorous in-flight calibration protocols and data analysis methodologies. 
These enable a thorough characterization of the instrumental performance and properties, laying a solid foundation for scientific discoveries. 
The in-orbit calibration of \textit{LEIA} also provides a robust paradigm for that of the WXT payload aboard the \textit{EP} satellite.
These calibration efforts and the resulting data serve as a valuable reference for future wide-field X-ray missions in time-domain astronomy.

\begin{acknowledgements}

We thank all the members of the EP team, the EP consortium, and the SATech-01 team. HQC thanks Dr. R. Nick Durham, Dr. Vadim Burwitz and Dr. Hui Sun for the comments. This work is supported by the Einstein Probe project, a space mission supported by Strategic Priority Program on Space Science of Chinese Academy of Sciences, in collaboration with ESA, MPE and CNES. The CAS team acknowledge contribution from ESA for calibration of the mirror assembly and tests of part of the devices. This work was partially supported by International Partnership Program of CAS (grant No. 113111KYSB20190020). The Leicester and MPE teams acknowledge funding by ESA. The work performed at MPE’s PANTER X-ray test facility has in part been supported by the European Union’s Horizon 2020 Program under the AHEAD2020 project (grant No. 871158). This work is supported by the National Science Foundation of China (grant nos. 12173055, 12173056, 12203071, 12203083). 

\end{acknowledgements}

%
%
\bibliography{leia-inorbit}
\bibliographystyle{aa}
\end{document}